\documentclass[aps,
               prd,
               superscriptaddress,
               amsfonts,
               amssymb,
               amsmath,
               preprintnumbers,
               nofootinbib,
               unsortedaddress,
               showpacs,
               tightenlines,
               floats,
               letterpaper,
               eqsecnum,
               ]{revtex4}

\usepackage{graphicx}
\usepackage{latexsym}
\usepackage{bm}
\usepackage[draft=false]{hyperref}
\usepackage[sort&compress]{natbib}

\def\C{{\mathcal C}}
\def\E{{\mathcal E}}
\def\F{{\mathcal F}}
\def\I{{\mathcal I}}

\def\O{{\mathcal O}}
\def\P{{\mathcal P}}
\def\R{{\mathcal R}}
\def\W{{\mathcal W}}

\begin{document}

\title{
Mass spectrum of primordial black holes from inflationary perturbation in the Randall-Sundrum braneworld: a limit on blue spectra
}

\author{Yuuiti Sendouda}\email[]{sendouda@utap.phys.s.u-tokyo.ac.jp}
\affiliation{
Department of Physics, Graduate School of Science, The University of Tokyo, 7-3-1 Hongo, Bunkyo-ku, Tokyo 113-0033, Japan
}
\author{Shigehiro Nagataki}
\affiliation{
Yukawa Institute for Theoretical Physics, Kyoto University, Kitashirakawa, Sakyo-ku, Kyoto 606-8502, Japan
}
\affiliation{
Kavli Institute for Particle Astrophysics and Cosmology, Stanford Linear Accelerator Center, 2575 Sand Hill Road, MS 29, Menlo Park, CA 94025, USA
}
\author{Katsuhiko Sato}
\affiliation{
Department of Physics, Graduate School of Science, The University of Tokyo, 7-3-1 Hongo, Bunkyo-ku, Tokyo 113-0033, Japan
}
\affiliation{
Research Center for the Early Universe, Graduate School of Science, The University of Tokyo, 7-3-1 Hongo, Bunkyo-ku, Tokyo 113-0033, Japan
}

\date{
\today
}

\begin{abstract}
The mass spectrum of the primordial black holes formed by density perturbation in the radiation-dominated era of the Randall-Sundrum type-2 cosmology is given.
The spectrum coincides with standard four-dimensional one on large scales but the deviation is apparent on smaller scales.
The mass spectrum is initially softer than standard four-dimensional one, while after accretion during the earliest era it becomes harder than that.
We also show expected extragalactic diffuse photon background spectra varying the initial perturbation power-law power spectrum and give constraints on the blue spectra and/or the reheating temperature.
The most recent observations on the small scale density perturbation from WMAP, SDSS and Lyman-$ \alpha $ Forest are used.
What we get are interpreted as constraints on the smaller-scale inflation on the brane connected to the larger one at the scale of Lyman-$ \alpha $ Forest.
If we set the bulk curvature radius to be $ 0.1~\text{mm} $ and assume the reheating temperature is higher than $ 10^6~\text{GeV} $, the scalar spectral index from the smaller scale inflation is constrained to be $ n \lesssim 1.3 $.
Typically, the constraints are tighter than the standard four-dimensional one, which is also revised by us using the most recent observations.
\end{abstract}



\preprint{UTAP-552}
\preprint{RESCEU-1/06}
\preprint{YITP-06-14}

\maketitle

\section{Introduction}

It is well believed that the life of heavy stars ends with luminous explosion collapsing into a black hole.
The birth of such massive black holes are daily phenomena in that we can observe supernovae at the present time.
On the other hand, the mechanisms to produce holes smaller than stellar mass in the early Universe had been invented several decades ago \cite{1967SvA....10..602Z,Hawking:1971ei}.
Those very small holes, called primordial black holes, are thought to be formed from density perturbation in the form of radiation, and their weight is comparable to the horizon mass in the radiation dominated era \cite{Carr:1974nx,Carr:1975qj}.
The primordial holes can exist everywhere in the Universe.
Those in the dense regions might have clustered, being one of the candidates for the dark matter.
The significance given to the primordial holes was not only their existence itself, but also the possibility of the observation of the Hawking radiation \cite{Hawking:1974rv,Hawking:1974sw}.
The quantum theory on the black hole background predicts that black holes radiate particles obeying almost black-body spectrum \cite{Page:1976df}.
It also predicts that the stellar mass holes cannot produce particle fields significantly, whereas primordial black holes are likely to emit various heavy particles, some of which may reach to the earth as cosmic rays.
One of the earliest work in this respect was done in \cite{Page:1976wx}, which considered the possible contribution of PBHs to the extragalactic gamma-ray background.
Constraints on the PBH abundance from cosmic rays have been extensively investigated \cite{1990PhRvD..41.3052M,1991PhRvD..44..376M,1991ApJ...371..447M}.
Primordial black holes serving as the sources of cosmic rays or relic particles were also studied in the literature \cite{1976ApJ...206....8C,1987Natur.329..308M}.
A turning point for PBHs comes with the discovery of the cosmological inflation \cite{Sato:1980yn,Guth:1980zm}.
They predicts almost Gaussian, super-horizon scale density perturbations which are thought not only to serve as seeds of the structure formation in late time, but an origin of primordial black holes.
Then, in addition to the large scale structure like the cosmic microwave background (CMB) or the spatial distribution of the galaxies, PBHs became a probe for rather small scale structure in the inflationary early Universe \cite{Carr:1994ar,Green:1997sz,Kim:1996hr,Kim:1999iv,Bringmann:2001yp,Blais:2002gw,Rubin:2000dq,Rubin:2001yw,Barrau:2002ru,Bugaev:2002yt}.

After some decades, the study of cosmology is shifting toward higher-dimensional models called braneworld.
The idea of a membrane as our Universe has been inspired by the string theory.
A toy model for the fundamental theory suggested by Randall and Sundrum \cite{Randall:1999ee,Randall:1999vf} is one of those which successfully matches with standard big-bang cosmology in the late time, but deviates in the earliest epoch of the radiation dominated-era \cite{Cline:1999ts,Csaki:1999jh,Binetruy:1999ut,Binetruy:1999hy} (see, for example, \cite{Maartens:2003tw} for a review).
In the context of such modern cosmology, we should rediscover the utility of primordial black hole.
Particularly, PBHs in the Randall-Sundrum (RS) cosmology show very different features from those in standard four-dimensional scenario \cite{Guedens:2002km,Guedens:2002sd,Majumdar:2002mr,Clancy:2003zd}.
They covers all the importance, i.e., formation dynamics, accretion process, and the Hawking radiation.
Primordial black hole is an exceedingly good probe for the brane early Universe.
Possibilities for constraining the braneworld geometry, such as the bulk curvature radius, have been shown by investigating the cosmic-ray spectra \cite{Sendouda:2003dc,Sendouda:2004hz}.

In this paper, we give the connection between the curvature perturbation spectrum generated on the brane after brany or bulky inflation (for example, \cite{Maartens:1999hf,Himemoto:2000nd,Koyama:2003yz}) and the resulting mass spectrum of the primordial black holes.
In addition, we constrain the blue spectral index of the curvature perturbation using the observational data of the extragalactic diffuse photon background.
The cosmological parameters we use are based on the concordance model containing the cosmological constant and cold dark matter ($ \Lambda $CDM), which has been recently confirmed by the observation of cosmic microwave background (CMB) by Wilkinson Microwave Anisotropy Probe (WMAP) \cite{Bennett:2003bz,Spergel:2003cb}.
For cosmological parameters, we use First Year WMAP and Two Degree Field system (2dF) data for the density, Hubble parameter, age of the Universe, and so on.
On the other hand, for smaller-scale matter power spectrum, we use the result of \cite{Seljak:2004xh}, which combined WMAP data with Lyman-$ \alpha $ forest (Ly$\alpha$F) and Sloan Digital Sky Survey (SDSS).
With these recent observational data, we also reexamine the constraint on the perturbation spectrum in four dimensions \cite{Kim:1999iv}.

This paper is organized as follows.
In Sec.~\ref{sec:PS}, we review the basic framework to compute mass spectrum of gravitationally bounded objects like primordial black holes from density perturbations by the aid the Press-Schechter theory.
Then after describing the Randall-Sundrum braneworld in Sec.~\ref{sec:RS}, we show the mass spectrum of primordial black holes in the RS cosmology.
Finally in Sec.~\ref{sec:constraint}, we give constraints on brane inflation via the curvature perturbation spectrum and the reheating temperature.

\section{
Mass spectrum of PBHs in the RS braneworld
}

\subsection{\label{sec:PS}
Press-Schechter prescription for inflationary density perturbation
}

In this section we summarize how to derive the mass spectrum of PBHs in the framework described by Press and Schechter \cite{1974ApJ...187..425P}.
In the four-dimensional theory there has been much effort to obtain the mass spectrum of PBHs.
Since we are now focusing on possible differences emerging from different dimensionality, the mass spectrum is described in a manner such that the differences will be clarified.
We assume the bulk spacetime has five dimensions while the matter content is confined in a four-dimensional brane.
Although we will only consider the five-dimensional Randall-Sundrum model, the following arguments will not strongly depend on the detail of the model.

First we start from the comoving number density of density fluctuation which arose in the inflationary early Universe.
For the moment we do not care about the mechanism of the inflation in the braneworld, which would be eventually included in the power spectrum of scalar perturbation.
Now we focus on the quantities evaluated at some fixed time in the radiation-dominated era.
The fluctuation is, after smoothed by some suitable window function, at every point spherical with comoving radius $ R $, which is the scale of the window function.
On comoving hypersurfaces, density fluctuation at $ \mathbf x $ is characterized by $ \delta(\mathbf x) \equiv (\rho(\mathbf x)-\rho)/\rho $, where $ \rho $, without explicit spatial dependence, represents the background energy density.
$ \delta $ is averaged by a window function $ W $ to give a smoothed quantity
\begin{equation}
\delta_R(\mathbf x)
 = \frac{1}{V}
   \int W(|\mathbf x'-\mathbf x|/R) \delta(\mathbf x')\mathrm d^3\mathbf x',
\end{equation}
where $ V = \int W(|\mathbf x'-\mathbf x|/R) \mathrm d^3\mathbf x' $ is the comoving volume contained in the window function.
The spectrum of the smoothed density contrast is given through its Fourier transform, $ \tilde\delta_R \equiv (2\pi)^{3/2} \int e^{-\mathrm i \mathbf k \cdot \mathbf x} \delta_R \mathrm d^3\mathbf x $, by taking the ensemble average as $ \P_\delta = [k^3/(2 \pi^2)] \langle |\tilde\delta_R|^2 \rangle $.
$ \P_\delta $, assumed to be independent of the direction of $ \mathbf k $, gives the expression for the variance
\begin{equation}
\sigma_R^2
 = \langle \delta_R^2(\mathbf x) \rangle
 = \int_0^\infty \tilde W^2(kR) \P_\delta(k) \frac{\mathrm dk}{k},
\end{equation}
where $ \tilde W $ is the volume-normalized Fourier transform of $ W $.
Below we will work throughout with those smoothed quantities $ \delta_R $, $ \sigma_R $, and $ \P_\delta $, so the indication of smoothing, $ R $, will be dropped hereafter.
We introduce $ M = \rho V $, the mass contained in the window function.
The comoving number density of smoothed fluctuations whose comoving radius and amplitude range within $ R \in [R,R+\mathrm dR] $ and $ \delta \in [\delta,\delta+\mathrm d\delta] $, respectively, is written as
\begin{equation}
n(R,\delta)
\mathrm dR \mathrm d\delta
 = \frac{1}{V} F(R,\delta) \mathrm dR \mathrm d\delta,
\label{eq:flucdensity}
\end{equation}
where $ F \mathrm dR \mathrm d\delta $ is the fraction of the space.
We assume the Gaussian probability distribution of the density fluctuation
\begin{equation}
P(R,\delta)
 = \frac{2}{\sqrt{2 \pi} \sigma}
   \exp\left(-\frac{\delta^2}{2\sigma^2}\right),
\end{equation}
where the prefactor $ 2 $ is of the Press-Schechter prescription.
It gives the fraction $ F $ as
\begin{equation}
F(R,\delta)
 = \left|\frac{\partial P}{\partial R}\right|
 = \sqrt{\frac{2}{\pi}} \frac{|\sigma'|}{\sigma^2}
   \exp\left(-\frac{\delta^2}{2 \sigma^2}\right)
   \left|\frac{\delta^2}{\sigma^2}-1\right|,
\end{equation}
where a prime denotes derivative over the scale of smoothing, $ R $.

Next we describe the spectrum of density perturbation.
Usual inflationary models predict that curvature perturbation on comoving hypersurfaces, $ \R_\mathrm c $, is produced during accelerating expansion, and its power spectrum $ \P_{\R_\mathrm c} $ tends to obey power-law.
Cosmological perturbation theory tells us how such a perturbation evolves and what we get as observables.
Now we want to know the spectrum of density perturbation, which form PBHs when they cross inward the cosmological horizon.
In the contexts of higher-dimensional cosmology, there are certain points to be taken seriously.
We first notice that the presence of the extra dimensions may highly affect the gravitational dynamics of scalar perturbation on a brane.
The effects can be categorized into two classes;
one is locally determined by matter on the brane while the other is the non-local responses of the bulk geometry to the perturbation on the brane.
The local effect coming from non-standard form of matter source is reflected in the background cosmological expansion, which of the Randall-Sundrum cosmology we review in the next subsection.
To see how the latter affects, let us take a look at the evolution of scalar perturbation in the RS cosmology.
In the super-horizon regime, above introduced $ \R_\mathrm c $ coincides with another variable $ \zeta $, the gauge-invariant curvature perturbation on uniform-density hypersurfaces \cite{Wands:2000dp}.
Four-dimensional inflationary theories predict that these values are determined by the inflaton expectation-value fluctuation and are almost constant outside the cosmological horizon \cite{Liddle:2000cg}.
On the other hand, in the five-dimensional Randall-Sundrum cosmology, a non-adiabatic Kaluza-Klein mode contributes to $ \zeta $ and generally it varies with time \cite{Koyama:2000cc,Langlois:2000iu,Koyama:2001ct,Maartens:2003tw}, which can modify the resulting matter spectrum.
In addition, the initial value measured at the exit of horizon during inflation is also modified by the higher-dimensional effect \cite{Koyama:2004ap,Yoshiguchi:2004nm,Koyama:2005ek}.
We should include such non-local effects from the bulk space for the complete arguments, however, to solve perturbation in the extra dimensions is difficult and, in particular, discussions for determination of initial scalar perturbation \cite{Koyama:2004ap,Yoshiguchi:2004nm,Koyama:2005ek} seems not fully satisfactory.
Therefore, in this paper we put apart such non-local effects but focus on the local background change.
In the language used in \cite{Langlois:2000iu} for the RS cosmology, what we do is to set those contributions from anisotropic stress, dark radiation, and isocurvature originating from higher-dimensional nature to be zero;
nonetheless, we are still able to capture one of the essential features of higher-dimensional cosmology.
The constancy of curvature perturbation and its power spectrum in the super-horizon regime is recovered in this framework.
A relevant brane inflation model is, for instance, what was treated in \cite{Maartens:1999hf}.

Now let us derive and investigate the density perturbation spectrum.
At sufficiently late times, i.e., four-dimensional regime, the comoving curvature perturbation $ \R_\mathrm c $ effectively takes a role as gravitational potential and induces observable density perturbation, whose spectrum $ \P_\delta $ on large scales is given in the following manner \cite{Liddle:2000cg}
\begin{equation}
\P_\delta(k,t)
 = T^2(k) \left(\frac{2+2w}{5+3w}\right)^2 \left(\frac{k}{aH}\right)^4
   \P_{\R_\mathrm c},
\label{eq:evo_ps}
\end{equation}
where $ k \equiv 1/R $ is the comoving wave length, $ w \equiv p/\rho $ represents the equation of state of the era considered, and the transfer function $ T(k) $ expresses the effect of deformation to matter perturbation.
$ H $ is the Hubble and $ a $ is the scale factor normalized to be unity at the present.
The time when the above spectrum is evaluated can be arbitrarily chosen if only the Universe can be regarded as four dimensional then.
Here we set the time by the condition that the Hubble radius becomes equal to some length $ a/k_0 $.
Then the variance is expressed as
\begin{equation}
\sigma^2
 = \int_0^\infty \tilde W^2(kR) \P_\delta(k,t(aH=k_0)) \frac{\mathrm dk}{k}.
\end{equation}
Through the above relations, the observed density perturbation spectrum $ \P_{\delta}^{(\mathrm{obs})} $ can be utilized to constrain the curvature perturbation.
Parameterizing as
\begin{equation}
\P_{\R_\mathrm c}(k)
 = A_\mathrm s \left(\frac{k}{k_\mathrm{pivot}}\right)^{n_\mathrm s-1},
\end{equation}
a combined result from First Year WMAP, SDSS galaxies (SDSS-gal), and Ly$ \alpha $ Forest claims $ A_\mathrm s = 2.45 \times 10^{-9} $ and $ n_\mathrm s = 0.971 $ at the pivot scale $ k_\mathrm{pivot} = 0.05~\text{Mpc}^{-1} $ \cite{Seljak:2004xh}.
This parameterization is valid down to Ly$ \alpha $F scale.
Its corresponding wave number is a rather useful choice as the reference scale for our purpose, i.e., we now take $ k_0 = 0.9h~\text{Mpc}^{-1} $.
This length is in the radiation-domination regime but is much larger than the scale at which the cosmological evolution is possibly modified.
Since the scales we wish to consider are much smaller than those appeared in the observations mentioned above, we now assume a single power-law spectrum below the Ly$ \alpha $F scale reparameterized as
\begin{equation}
\P_{\R_\mathrm c}(k)
 = A \left(\frac{k}{k_0}\right)^{n-1},
\end{equation}
where the amplitude is $ A = A_\mathrm s (k_0/k_\mathrm{pivot})^{n_s-1} = 2.28 \times 10^{-9} $, but now that the spectral index $ n $ is regarded as a free parameter.
In other words, what is assumed here to realize such a broken power-law is some generation mechanism like the double inflation model where the connecting scale between the first and second inflations is placed at the Ly$ \alpha $ Forest scale.
From this point of view, what we will consider can be comprehensively interpreted as limits on the second inflation.
Although not all the brane inflation models prefer such a fluctuation spectrum, we should take it into account that an enhancement of small scale density perturbation is observationally inferred by the First Year WMAP result \cite{Bennett:2003bz,Spergel:2003cb}.

Substituting the spectrum into Eq.~(\ref{eq:evo_ps}) and setting $ w = 1/3 $ give the explicit form for the variance as
\begin{equation}
\sigma^2
 = \left(\frac{4}{9}\right)^2 \W^2 A (k_0 R)^{-(3+n)}
 \equiv \sigma_0^2 (k_0 R)^{-(3+n)},
\end{equation}
where we defined $ \W $ as
\begin{equation}
\W^2(R)
  \equiv \int_0^\infty \tilde W^2(x) T^2(x/R) x^{2+n} \mathrm dx,
\end{equation}
and $ \sigma_0^2 \equiv (4/9)^2 \W^2 A $.
If we choose the Gaussian filter $ \tilde W = e^{-k^2 R^2/2} $ and, for simplicity, assume $ T = 1 $, then one gets $ \W^2 = (1/2) \Gamma[(3+n)/2] $.
The volume inside the Gaussian filter is $ V = (\sqrt{2 \pi} R)^3 $.

Then we consider gravitational collapse occurring at the horizon entry.
We denote the corresponding time as $ t_\mathrm h $.
We introduce ``linear evolution'' factor $ D $ by which the amplitude and root-mean-square of density perturbation crossing inward the horizon are related to the values at the reference scale $ k_0 $ as\footnote{
The time sequence $ t(aH=k_0) > t_\mathrm h $ is understood by the negative power of $ D $.
}
\begin{equation}
\left\{
\begin{matrix} \delta_\mathrm h \\ \sigma_\mathrm h \end{matrix}
\right\}
\equiv D^{-1}
\left\{
\begin{matrix} \delta \\ \sigma \end{matrix}
\right\}.
\label{eq:evo_delta}
\end{equation}
The following discussion does not depend on the detailed form of $ D $, which we will give in the next subsection.
The collapsed black hole mass, $ M_\mathrm{bh} $, is determined by the amplitude $ \delta_\mathrm h $ and the comoving scale $ R $.
Then the comoving mass spectrum of primordial black holes formed by density perturbation is now given as
\begin{equation}
\frac{\mathrm dn_\mathrm{bh}}{\mathrm dM_\mathrm{bh}}
 = \int_{\delta_{\mathrm h,\mathrm{min}}}^{\delta_{\mathrm h,\mathrm{max}}}
   n(R,\delta) J(M_\mathrm{bh},\delta_\mathrm h) \mathrm d\delta_\mathrm h,
\end{equation}
where the domain of integration is determined for each collapse model and
\begin{equation}
J(M_\mathrm{bh},\delta_\mathrm h)
 \equiv \frac{\partial (R,\delta)}{\partial (M_\mathrm{bh},\delta_\mathrm h)}
\end{equation}
is the Jacobian matrix.
The integration is performed with fixed $ M_\mathrm{bh} $.
Substituting Eq.~(\ref{eq:flucdensity}) and defining $ \mathrm dx \equiv \mathrm d\delta_\mathrm h/\sigma_\mathrm h $, the above mass spectrum becomes
\begin{equation}
\frac{\mathrm dn_\mathrm{bh}}{\mathrm dM_\mathrm{bh}}
 = \sqrt{\frac{2}{\pi}}
   \int_{x_\mathrm{min}}^{x_\mathrm{max}}
   \frac{J}{V D}
   \frac{|\sigma'|}{\sigma}
   e^{-x^2/2} |x^2-1| \mathrm dx.
\end{equation}
In usual cases, the scale $ R $ determines the time of horizon entry $ t_\mathrm h $.
This deduces the functional form $ \delta_\mathrm h $ to be
\begin{equation}
\delta_\mathrm h = D(R)^{-1} \delta.
\end{equation}
The collapsed mass $ M_\mathrm{bh} $ may be a function of both the comoving scale $ R $ and the amplitude $ \delta_\mathrm h $.
In ordinary parameterization, the form of $ M_\mathrm{bh} $ can be expressed as
\begin{equation}
M_\mathrm{bh} = f(\delta_\mathrm h) M_\mathrm h(R),
\end{equation}
where $ M_\mathrm h $ is the mass contained in the Hubble volume, the horizon mass, now being evaluated at the time of horizon entry, and $ f $ is the fraction.
Then the Jacobian is found to be $ J = D/M_\mathrm{bh}' $ and the formula for the mass spectrum reduces to
\begin{equation}
\frac{\mathrm dn_\mathrm{bh}}{\mathrm dM_\mathrm{bh}}
 = \sqrt{\frac{2}{\pi}}
     \int_{x_\mathrm{min}}^{x_\mathrm{max}}
     \frac{1}{V M_\mathrm{bh}'}
     \frac{|\sigma'|}{\sigma}
     e^{-x^2/2} |x^2-1| \mathrm dx.
\label{eq:ms}
\end{equation}
Gravitational collapse taking place at the horizon-entrance is an issue even in four dimensions, for instance, some critical phenomenon has been found \cite{Niemeyer:1999ak}.
The process may also be affected by dimensionality.
One simple prescription is to regard the emerging black hole mass almost the same as the horizon mass at the horizon entrance, i.e., $ f = \text{const} \sim 1 $.
Actually Carr argued that $ f = w^{3/2} $ where $ w \equiv p/\rho $ \cite{Carr:1975qj}.
For this reason, we here assume that $ f $ is almost constant and will concentrate on the relation between the comoving scale and horizon mass.
Four dimensional constraints from the critical collapse model have been considered elsewhere \cite{Niemeyer:1997mt,Yokoyama:1998xd}.

We estimate the PBH mass spectrum a bit more assuming constant $ f $.
Another assumption that both $ M_\mathrm{bh} $ and $ \sigma $ are power functions of $ R $ gives $ |\sigma'|/\sigma V M_\mathrm{bh}' \sim 1/V M_\mathrm{bh} $, so that one obtains
\begin{equation}
\frac{\mathrm dn_\mathrm{bh}}{\mathrm dM_\mathrm{bh}}
 \sim \frac{1}{V M_\mathrm{bh}}
      \int_{x_\mathrm{min}}^{x_\mathrm{max}} e^{-x^2/2} |x^2-1| \mathrm dx
 \equiv \frac{1}{V M_\mathrm{bh}} \I(M_\mathrm{bh}).
\end{equation}
The integration is performed to give
\begin{equation}
\I
 =
\begin{cases}
I(x_\mathrm{min})-I(x_\mathrm{max}) & x_\mathrm{min} \geq 1 \\
I(x_\mathrm{max})-I(x_\mathrm{min}) & x_\mathrm{max} \leq 1 \\
2 I(1)-I(x_\mathrm{max})-I(x_\mathrm{min}) & \text{otherwise}
\end{cases},
\end{equation}
where $ I(x) \equiv x e^{-x^2/2} $.
Now we observe that the denominator $ V M_\mathrm{bh} $ gives the ``scale-invariant'' part of the mass spectrum that does not depend on the density perturbation spectrum.
On the other hand, the integral $ \I $ determines its scale-dependent tilt through $ \sigma_\mathrm h $.
In general $ \delta_{\mathrm h,\mathrm{min}} \sim \delta_{\mathrm h,\mathrm{max}} \sim \O(1) $.
If we adopt power-law $ \sigma_\mathrm h \propto k^\alpha $ normalized on the large scale $ k_0^{-1} $ as $ \sigma_\mathrm h \sim \sqrt A \sim \O(10^{-5}) $, then we deduce that the PBH abundance never possesses finite values unless $ \sigma_\mathrm h $ at least grows up to $ \delta_{\mathrm h,\mathrm{min}} \sim \O(1) $.
Hence, hereafter we consider such blue spectrum cases with positive $ \alpha $, where the explicit form of $ \alpha $ expressed with spectral index $ n $ will be found later.
In such a situation, there would be a threshold scale $ R_\mathrm{th} $ at which $ \sigma_\mathrm h \sim \delta_{\mathrm h,\mathrm{min}} $.
$ \I $ is estimated as $ \I \simeq x_{\mathrm{min}} e^{-x_{\mathrm{min}}^2/2} $ above the threshold, while $ \I \simeq x_{\mathrm{max}} $ below it.
Therefore the approximated expression for the tilt is given as
\begin{equation}
\I
 \simeq \sigma_\mathrm h^{-1}(R)
 \times
\begin{cases}
\delta_\mathrm{h,min}
\exp\left[-\frac{\delta_\mathrm{h,min}^2}{2 \sigma_\mathrm h^2(R)}\right]
 & R \gtrsim R_\mathrm{th} \\
\delta_\mathrm{h,max}
 & R \lesssim R_\mathrm{th} \\
\end{cases}.
\end{equation}

We can also evaluate the mass density of the PBHs from the mass spectrum which has been often used in the literature.
$ \rho_{\mathrm{bh}}(>M) $, the mass density of black holes whose mass is greater than $ M $, can be evaluated as
\begin{equation}
\rho_{\mathrm{bh}}(>M)
 = \int_M^\infty M_\mathrm{bh}
                \frac{\mathrm dn_{\mathrm{bh}}}{\mathrm dM_\mathrm{bh}}
                \mathrm dM_\mathrm{bh}
 \sim \int_M^\infty \frac{\I(M_\mathrm{bh})}{V(M_\mathrm{bh})}
      \mathrm dM_\mathrm{bh},
\end{equation}
which further leads to
\begin{equation}
\rho_{\mathrm{bh}}(>M)
 \sim \frac{M}{V(M)}
 \times
\begin{cases}
x_{\mathrm{min}}(M)^{-1} e^{-x_{\mathrm{min}}(M)^2/2}
 & M \gtrsim M_\mathrm{th} \\
x_{\mathrm{max}}(M)
 & M \lesssim M_\mathrm{th}
\end{cases},
\end{equation}
where $ M_\mathrm{th} $ is the mass scale associated with $ R_\mathrm{th} $.
This $ \rho_\mathrm{bh} $ is dominated by the holes whose mass is $ M $ because the mass spectrum is a rapidly decreasing function with increasing $ M_\mathrm{bh} $.
It is useful to define the mass fraction of black holes relative to the background.
The fraction of black holes whose mass is (larger than) $ M $ is given as $ \beta(M) \equiv \rho_{\mathrm{bh}}(>M)/\rho $.
This is a time-dependent quantity.
For convenience, we use $ \beta_\mathrm i $, which is the ratio to the background energy density at the time of the formation of holes with mass $ M $.
Since such a radiation energy density is given by $ \rho_\mathrm i = M/V $, the initial fraction is obtained as
\begin{equation}
\beta_\mathrm i(M)
 \sim
\begin{cases}
\sigma_\mathrm h(M)
\exp\left[-\frac{\delta_{\mathrm h,\mathrm{min}}^2}
                {2 \sigma_\mathrm h(M)^2}\right]
 & M \gtrsim M_{\mathrm{th}} \\
\sigma_\mathrm h(M)^{-1}
 & M \lesssim M_{\mathrm{th}}
\end{cases}.
\end{equation}
We note that with the above assumptions this $ \beta_\mathrm i $ is equivalent to $ \alpha_\mathrm i $ defined as the ratio of mass density of holes relative to radiation energy density at the time of formation, which has been also used in the past works \cite{Green:1997sz,Clancy:2003zd,Sendouda:2003dc,Sendouda:2004hz,Tikhomirov:2005bt}.

In the following our task is to obtain expressions for $ V $, $ M_\mathrm{bh} $, and $ \sigma_\mathrm h $ according to each cosmological model.
We then will be able to constrain inflationary models through the perturbation spectrum by constraints on the PBH abundance.

\subsection{\label{sec:RS}
Randall-Sundrum braneworld cosmology
}

In this section we describe the background cosmology and the evolution of density perturbation in the so-called Randall-Sundrum type-2 braneworld \cite{Randall:1999ee,Randall:1999vf}.
This theory introduces small-scale modification whereas the late-time behavior converges to the purely four dimensional one \cite{Cline:1999ts,Csaki:1999jh,Binetruy:1999ut,Binetruy:1999hy}.
Hereafter we basically work with the units such that all the four-dimensional Planck units are unity: $ M_4 = T_4 = \ell_4 = t_4 = 1 $, although sometimes they appear in expressions.

The RS2 braneworld is composed by a four dimensional timelike brane, which has positive tension, embedded in five dimensional anti de Sitter (AdS) bulk.
Standard model matter is confined within the brane.
The cosmological expansion on the brane is determined by the modified Friedmann equation
\begin{equation}
H^2
 = \frac{8\pi}{3 M_4^2}
   \left[\rho \left(1+\frac{\rho}{2\lambda}\right)+\rho_\E\right]
 + \frac{\Lambda_4}{3}
 - \frac{K}{a^2},
\label{eq:Friedmann_modified}
\end{equation}
where $ \lambda > 0 $ is the tension of the brane, $ \rho_\E $ the energy density of the so-called dark radiation, and $ K $ the topology of the three dimensional space.
The effective Planck scale $ M_4 $ and effective cosmological constant on the brane $ \Lambda_4 $ are, respectively, given as
\begin{equation}
M_4^2
 = \frac{3}{4\pi}\frac{M_5^6}{\lambda},
\Lambda_4
 = \frac{4\pi}{M_5^3}
   \left(\frac{4\pi}{3}\frac{\lambda^2}{M_5^3}-|\Lambda_5|\right),
\end{equation}
where $ M_5 $ is the fundamental mass scale in five dimensions and $ \Lambda_5 < 0 $ is the cosmological constant in the bulk.
It is useful to define the scale of the theory by the curvature radius of bulk AdS $ \ell \equiv \sqrt{3 M_5^3/(4 \pi |\Lambda_5|)} $.
Setting $ \Lambda_4 $ sufficiently small, the relation between the geometrical scale and the brane tension is obtained as $ \lambda = 3 M_4^2/(4\pi \ell^2) $.
Then we get $ M_4^2 = \ell M_5^3 $.
In the following we omit $ \Lambda_4 $ and $ K $, which have been confirmed to be sufficiently small by WMAP, and also drop $ \rho_\E $ \cite{Ichiki:2002eh}.
Then it is obtained that the cosmological expansion in the radiation dominated era is separated into two parts, namely in the earlier one the $ \rho^2/\lambda $ term dominates (``$ \rho $-square era'') whereas in the latter the $ \rho $ term does.
We often denote the two separated phases simply by ``$ \rho^2 $'' and ``$ \rho^1 $'', respectively.
For later use, we define the effective energy density and pressure
\begin{align}
\rho_\mathrm{eff}
 & = \rho \left(1+\frac{\rho}{2\lambda}\right), \\
p_\mathrm{eff}
 & = p + \frac{\rho}{2\lambda} (2p+\rho),
\end{align}
which give the modified Friedmann equation Eq.~(\ref{eq:Friedmann_modified}).
We can represent the effective equation of state as
\begin{equation}
w_\mathrm{eff}
 \equiv \frac{p_\mathrm{eff}}{\rho_\mathrm{eff}}
 =
\begin{cases}
\frac{5}{3} & \rho^2 \\
\frac{1}{3} & \rho^1
\end{cases}.
\end{equation}
Note that in our formulation, all the expressions and corresponding results associated with the $ \rho^1 $ case completely agree with the standard four-dimensional cosmology since the $ \ell \to 0 $ limit recovers the 4D cosmology.

The solution through the whole radiation-dominated era is
\begin{equation}
a(t)
 = a_\mathrm{eq} \frac{t^{1/4} (t+t_\mathrm c)^{1/4}}{t_\mathrm{eq}^{1/2}},
\end{equation}
where $ t_\mathrm c \equiv \ell/2 $ and the normalization is given at the time of matter-radiation equality $ t_\mathrm{eq} $.
The Hubble parameter is given as
\begin{equation}
H(t)
 = \frac{2t+t_\mathrm c}{4t (t+t_\mathrm c)}.
\end{equation}
Other background quantities, radiation energy density $ \rho $ and the horizon mass $ M_\mathrm h \equiv 4 \pi \rho/(3 H^3) $, are calculated as
\begin{align}
\rho(t)
 & = \frac{3}{4 \pi \ell^2}
     \left[\sqrt{H^2 \ell^2+1}-1\right]
   = \frac{3}{32 \pi t (t+t_\mathrm c)}, \\
M_\mathrm h(t)
 & = \frac{1}{H^3 \ell^2}
     \left[\sqrt{H^2 \ell^2+1}-1\right]
   = \frac{8 t^2 (t+t_\mathrm c)^2}{(2t+t_\mathrm c)^3}.
\end{align}
It is useful to note the asymptotic behavior in respective regime: $ M_\mathrm h \propto t^2 $ for $ \rho^2 $ and $ M_\mathrm h \propto t $ for $ \rho^1 $.
Knowing the background cosmological evolution, the horizon-crossing time $ t_\mathrm h $ for a density fluctuation with radius $ R = 1/k $ is determined by a condition
\begin{equation}
R a(t_\mathrm h) = H^{-1}(t_\mathrm h)
\end{equation}
Now let us derive the connection between the quantities at some fixed time $ (R,\delta) $ and those at the time of gravitational collapse $ (M_\mathrm h,\delta_\mathrm h) $.
The comoving radius $ R $ determines the horizon crossing time $ t_\mathrm h $ from the condition
\begin{equation}
R
 = \frac{t_\mathrm{eq}^{1/2}}{a_\mathrm{eq}}
   \frac{4 t_\mathrm h^{3/4} (t_\mathrm h+t_\mathrm c)^{3/4}}
        {2t_\mathrm h+t_\mathrm c},
\end{equation}
while the horizon mass at $ t_\mathrm h $ is evaluated as $ M_\mathrm h(t_\mathrm h) $.
Eliminating $ t_\mathrm h $ from these expressions gives
\begin{equation}
k_0 R
 \simeq
\begin{cases}
\left(\frac{M_\mathrm h^{3/4} M_\mathrm c^{1/4}}{M_{\mathrm h,0}}\right)^{1/2}
 & \rho^2 \\
\left(\frac{M_\mathrm h}{M_{\mathrm h,0}}\right)^{1/2}
 & \rho^1
\end{cases},
\end{equation}
where $ M_{\mathrm h,0} \equiv M_\mathrm h[t(aH=k_0)] \approx 1.94 \times 10^{47}~\text{g} $ is the horizon mass at our normalization scale and we defined $ M_\mathrm c \equiv \ell/4 \approx 3.37 \times 10^{25} (\ell/0.1~\text{mm})~\text{g} $.
One can adopt the above relation to the primordial power-law perturbation spectrum to find
\begin{align}
\sigma
 & \simeq \sigma_0 \times
\begin{cases}
\left(\frac{M_\mathrm h^{3/4} M_\mathrm c^{1/4}}
           {M_{\mathrm h,0}}\right)^{-(3+n)/4}
 & \rho^2 \\
\left(\frac{M_\mathrm h}{M_{\mathrm h,0}}\right)^{-(3+n)/4}
 & \rho^1
\end{cases}.
\end{align}

Next, we consider the evolution of scalar perturbations.
We focus on super-horizon regime and, for the reasons described in the previous subsection, omit anisotropic stress, dark radiation including its perturbation, and isocurvature.
Then it follows that the gauge-invariant curvature perturbation $ \zeta $ conserves \cite{Koyama:2000cc,Langlois:2000iu,Koyama:2001ct}, which leads to, in the longitudinal gauge \cite{Kodama:1985bj}, metric perturbation $ \Phi $ (now equals to $ -\Psi $) is also constant \cite{Langlois:2000iu}:
\begin{equation}
\Phi
 = \frac{3+3w_\mathrm{eff}}{5+3w_\mathrm{eff}} \zeta.
\end{equation}
On large scales $ \zeta $ coincides with $ \R_\mathrm c $.
From the remaining perturbed Einstein equation, the relation between $ \Phi $ and $ \delta\rho $ to the linear order is found as
\begin{equation}
\frac{4 \pi}{M_4^2} \left(1+\frac{\rho}{\lambda}\right) \delta\rho
 + a^{-2} \nabla^2\Phi = 0,
\end{equation}
where a nabla denotes vector derivatives over the comoving spatial coordinates.
From the above relations, we eventually find the time evolution of $ \delta $, or equivalently that of its Fourier transform $ \tilde\delta $, as
\begin{equation}
\tilde\delta
 = 
\begin{cases}
\frac{4}{15} \left(\frac{k}{aH}\right)^2 \tilde\R_\mathrm c & \rho^2 \\
\frac{4}{9} \left(\frac{k}{aH}\right)^2 \tilde\R_\mathrm c & \rho^1
\end{cases}.
\end{equation}
The latter was used to give Eq.~(\ref{eq:evo_ps}).
The evolution factor $ D $ is then obtained as
\begin{equation}
D^{-1}
 = \frac{\sigma_\mathrm h}{\sigma}
 = \frac{\delta_\mathrm h}{\delta}
 \simeq
\begin{cases}
\frac{3 M_\mathrm h^{3/4} M_\mathrm c^{1/4}}{5 M_{\mathrm h,0}}
 & \rho^2 \\
\frac{M_\mathrm h}{M_{\mathrm h,0}}
 & \rho^1
\end{cases}.
\end{equation}
Finally we get
\begin{align}
\sigma_\mathrm h
 \simeq \sigma_0 \times
\begin{cases}
\frac{3}{5} \left(\frac{M_\mathrm h^{3/4} M_\mathrm c^{1/4}}
                       {M_{\mathrm h,0}}\right)^{(1-n)/4}
 & \rho^2 \\
\left(\frac{M_\mathrm h}{M_{\mathrm h,0}}\right)^{(1-n)/4}
 & \rho^1
\end{cases}.
\end{align}

Throughout the above arguments, we have implicitly assumed that the beginning of the radiation-dominated era is before the transition.
Otherwise we do not see the effect of the fifth dimension.
The time of the beginning is regarded as the end of inflation, namely, reheating.
The reheating time $ t_\mathrm{rh} $ is expected to satisfy $ t_\mathrm{rh} < t_\mathrm c $.
The scale associated with reheating will be discussed in the next subsection together with the PBH mass spectrum.

\subsection{\label{sec:massspectrum}
The mass spectrum of PBH
}

\begin{figure}[htb]
\begin{center}
\includegraphics{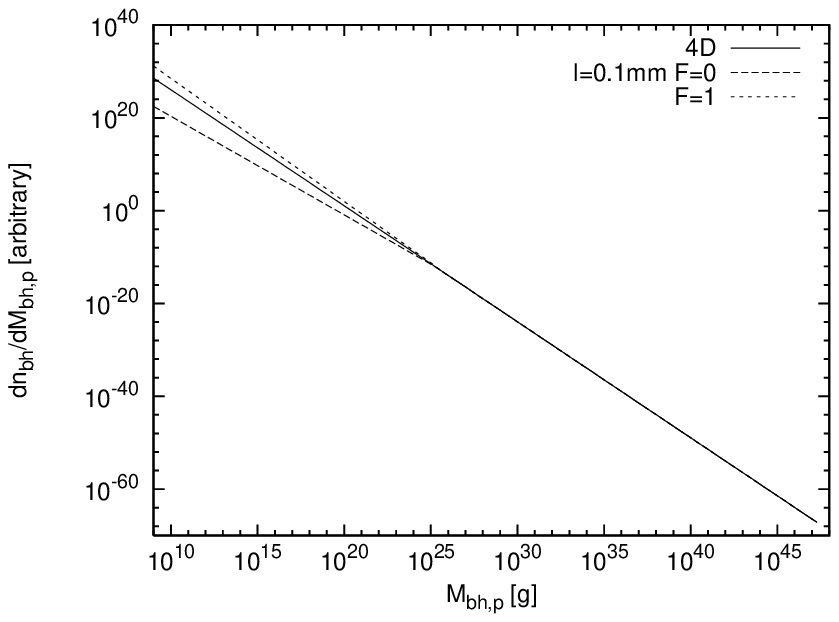}
\includegraphics{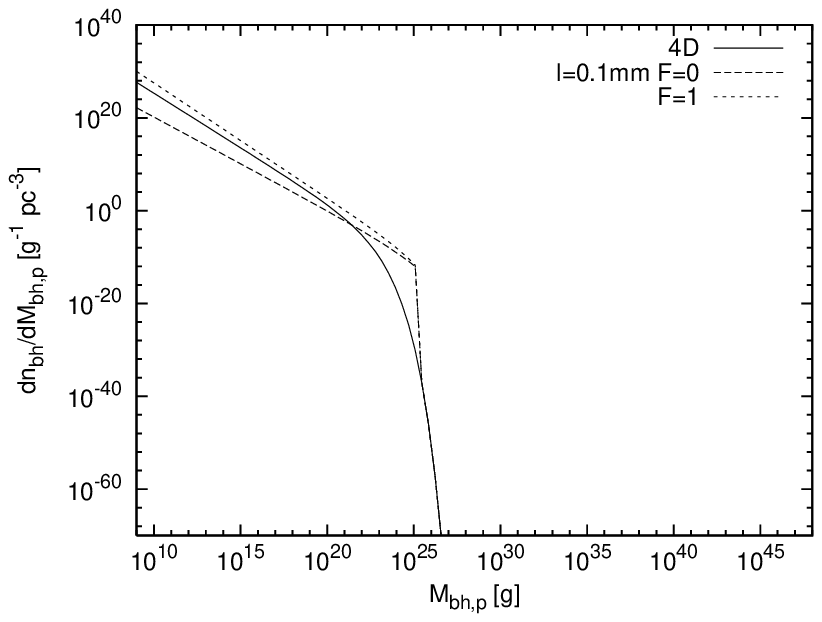}
\caption{
Left panel: $ n = 1.00 $ case, or the scale invariant part, of the mass spectrum.
The spectral index for standard 4D is $ -5/2 $.
5D spectra coincide with it above the threshold $ f M_\mathrm c \sim 10^{26}~\text{g} $, whereas below that, their indices deviate from $ -5/2 $: $ -17/8 = -2.125 $ for $ F = 0 $ and $ -2.65 $ for $ F = 1 $, respectively.
$ F = 0 $ spectrum is also regarded as the {\em initial} mass spectrum before accretion.
Values on the vertical axis are only expressing the relative scale.
Right panel: $ n = 1.60 $ case.
Heavier holes than $ M_\mathrm{th} $ are exponentially suppressed.
}
\label{fig:ms}
\end{center}
\end{figure}

Now we have obtained all the relations connecting the initial PBH mass $ M_\mathrm{bh} $ and comoving scales such as $ R $ through the parameters specifying cosmology, i.e., in this case $ \ell $.
Here the word {\em initial} means the spectrum is given as a function of the mass soon after gravitational collapse.
In this sense, we denote it as $ M_{\mathrm{bh},\mathrm i} $.
Again we set $ f $ to be constant, then the mass spectrum is written down as
\begin{equation}
\frac{\mathrm dn_\mathrm{bh}}{\mathrm dM_{\mathrm{bh},\mathrm i}}
 = \sqrt{\frac{2}{\pi}}
   \frac{3+n}{4}
   \frac{\rho_\mathrm{eq} M_\mathrm{eq}^{1/2}}{(1+z_\mathrm{eq})^3}
   \frac{\delta_{\mathrm h,\mathrm{min}} e^{-\frac{\delta_\mathrm{h,min}^2}
                {2 \sigma_\mathrm h^2(M_{\mathrm{bh},\mathrm i})}}}
        {\sigma_\mathrm h(M_{\mathrm{bh},\mathrm i})}
 \times
\begin{cases}
\frac{3}{4} f^{9/8} M_\mathrm c^{-3/8} M_{\mathrm{bh},\mathrm i}^{-17/8}
 & \rho^2 (M_{\mathrm{bh},\mathrm i} \lesssim f M_\mathrm c) \\
f^{3/2} M_{\mathrm{bh},\mathrm i}^{-5/2}
 & \rho^1 (M_{\mathrm{bh},\mathrm i} \gtrsim f M_\mathrm c)
\end{cases},
\label{eq:ms1}
\end{equation}
where
\begin{equation}
\sigma_\mathrm h^{-1}
 = \frac{9}{4 \W \sqrt A}
   \times
\begin{cases}
\frac{5}{3} \left(\frac{M_{\mathrm{bh},\mathrm i}^{3/4} M_\mathrm c^{1/4}}
                       {f^{3/4} M_{\mathrm h,0}}\right)^{(n-1)/4} 
 & \rho^2 (M_{\mathrm{bh},\mathrm i} \lesssim f M_\mathrm c) \\
\left(\frac{M_{\mathrm{bh},\mathrm i}}{f M_{\mathrm h,0}}\right)^{(n-1)/4}
 & \rho^1 (M_{\mathrm{bh},\mathrm i} \gtrsim f M_\mathrm c)
\end{cases}.
\end{equation}
We defined $ \rho_\mathrm{eq} \equiv \rho(t_\mathrm{eq}) $ and $ M_\mathrm{eq} \equiv \rho_\mathrm{eq} V(R_\mathrm{eq})/(1+z_\mathrm{eq})^3 \sim M_\mathrm h(t_\mathrm{eq}) \sim 10^{51}~\text{g} $.
For $ \delta_{\mathrm h,\mathrm{min}} $, we refer to \cite{Kawasaki:2004jd}, in which the collapse threshold on the $ \rho^2 $ background was found to be $ \delta_{\mathrm h,\mathrm{min}} \approx 0.1 $ rather than $ 1/3 $ of the ordinary radiation dominated era \cite{Carr:1975qj}.
The mass spectrum coincides with the four-dimensional one at larger scales, while on smaller scales it is deformed to be soft.
In the RS cosmology, we further incorporate an additional effect on the evolution of PBHs, which is {\em not} the mass loss via Hawking radiation.
On the contrary to the ordinary 4D case, radiation fluid accretes onto PBHs efficiently in the $ \rho $-square phase due to the slow expansion rate, which causes power-law growth of the mass $ M_\mathrm{bh}(t) \propto t^{2F/\pi} $ \cite{Majumdar:2002mr,Guedens:2002sd}, where $ 0 \leq F \lesssim 1 $ is the efficiency of accretion (for detailed treatment of accretion and its consequences, see \cite{Tikhomirov:2005bt}).
Only smaller holes formed in the $ \rho $-square phase suffer accretion soon after their formation till the end of the phase.
Since PBHs now being interested in are those which survive until at least the epoch of nucleosynthesis, we can ignore their mass loss during $ \rho $-square phase.
Then the mass at the end of accretion, which we denote {\em primordial} mass $ M_{\mathrm{bh},\mathrm p} $, is given as
\begin{equation}
M_{\mathrm{bh},\mathrm p}
 = \left(\frac{t_\mathrm c}{t_\mathrm h}\right)^{2F/\pi} M_{\mathrm{bh},\mathrm i}
 \simeq (16 f M_\mathrm c)^{F/\pi} M_{\mathrm{bh},\mathrm i}^{(\pi-F)/\pi}.
\end{equation}
Inverting this gives $ M_{\mathrm{bh},\mathrm i} = (16 f M_\mathrm c)^{-F/(\pi-F)} M_{\mathrm{bh},\mathrm p}^{\pi/(\pi-F)} $.
The PBH mass spectrum with respect to the primordial mass $ M_{\mathrm{bh},\mathrm p} $ is given by
\begin{align}
\frac{\mathrm dn_\mathrm{bh}}{\mathrm dM_{\mathrm{bh},\mathrm p}}
 & = \frac{\mathrm dM_{\mathrm{bh},\mathrm i}}{\mathrm dM_{\mathrm{bh},\mathrm p}}
     \frac{\mathrm dn_\mathrm{bh}}{\mathrm dM_{\mathrm{bh},\mathrm i}} \\
 & = \sqrt{\frac{2}{\pi}}
   \frac{3+n}{4}
   \frac{\rho_\mathrm{eq} M_\mathrm{eq}^{1/2}}{(1+z_\mathrm{eq})^3}
   \frac{\delta_{\mathrm h,\mathrm{min}} e^{-\frac{\delta_\mathrm{h,min}^2}
                {2 \sigma_\mathrm h^2(M_{\mathrm{bh},\mathrm i})}}}
        {\sigma_\mathrm h(M_{\mathrm{bh},\mathrm i})}
 \times
\begin{cases}
\frac{3}{4} \left(1+\frac{8}{9}\F\right)
16^\F f^{9/8+\F} M_\mathrm c^{-3/8+\F} M_{\mathrm{bh},\mathrm p}^{-17/8-\F}
 & \rho^2 (M_{\mathrm{bh},\mathrm p} \lesssim f M_\mathrm c) \\
f^{3/2} M_{\mathrm{bh},\mathrm p}^{-5/2}
 & \rho^1 (M_{\mathrm{bh},\mathrm p} \gtrsim f M_\mathrm c)
\end{cases}
\label{eq:ms2}
\end{align}
where $ \F \equiv 9F/[8(\pi-F)] $.
Mass spectra for $ \ell = 0.1~\text{mm} $ and $ f = 1 $ are shown in Fig.~\ref{fig:ms}.
The left panel shows $ n = 1.00 $ case (scale-invariant), whereas the right is for the $ n = 1.60 $ case.
In each panel, accretion efficiency for two 5D spectra is taken to be $ F = 1 $ and $ F = 0 $.

We note that the scale of reheating sets the small-scale cut-off in the spectrum.
In our treatment, the horizon mass at the reheating $ M_{\mathrm h,\mathrm{rh}} \equiv M_\mathrm h(t_\mathrm{rh}) $ gives the minimum mass in the spectrum.
The initial condition after inflation is comprehensively given via the reheating temperature $ T_\mathrm{rh} $.
In the $ \rho $-square era the modified Friedmann equation gives $ H^2 \propto \rho^2/\lambda \propto g^2 T^8 \ell^2 $, where we used the ordinary expression for radiation energy density $ \rho = (\pi^2/30) g T^4 $ with $ g $ the particle degree of freedom.
Hence horizon mass at the reheating time is
\begin{align}
M_{\mathrm h,\mathrm{rh}}
 & = \left(\frac{45}{2 \pi^3}\right)^2 g^{-2} \ell^{-3} T_{\mathrm{rh}}^{-8} \\
 & = 2.37 \times 10^3 \left(\frac{g}{100}\right)^{-2}
     \left(\frac{\ell}{0.1~\text{mm}}\right)^{-3}
     \left(\frac{T_{\mathrm{rh}}}{10^6~\text{GeV}}\right)^{-8} ~\text{g},
\label{eq:mass_rh}
\end{align}
which corresponds to the smallest PBH mass $ M_{\mathrm{bh},\mathrm{min}} = f M_{\mathrm h,\mathrm{rh}} $.
If one wants the Universe to begin in the $ \rho^2 $-era, then $ M_{\mathrm h,\mathrm{rh}} \lesssim M_\mathrm c $ is required.
This claims the minimum reheating temperature for the existence of 5D PBHs
\begin{equation}
T_\mathrm{rh}
  \gtrsim 10^3 \left(\frac{\ell}{0.1~\text{mm}}\right)^{-1/4}~\text{GeV}.
\label{eq:T_rhmin}
\end{equation}
In the 4D case, $ H^2 \propto \rho \propto g T^4 $, so
\begin{align}
M_{\mathrm h,\mathrm{rh}}
 & = \sqrt{\frac{45}{16\pi^3}} g^{-1/2} T_{\mathrm{rh}}^{-2} \\
 & = 9.75 \times 10^3 \left(\frac{g}{100}\right)^{-1/2}
     \left(\frac{T_{\mathrm{rh}}}{10^{14}~\text{GeV}}\right)^{-2} ~\text{g}.
\end{align}

\section{\label{sec:constraint}
Constraint on the perturbation spectrum from PBHs
}

We are interested in giving constraints on the density perturbation using limits on the PBH abundance.
In our case, it will lead to informations on inflation in the RS braneworld.
We use the large-scale normalization $ \sigma_0 $ given through $ A_\mathrm s $ obtained from WMAP+SDSS-gal+Ly$ \alpha $F \cite{Seljak:2004xh}, and assume that the power spectrum below the Ly$ \alpha $F normalization scale obeys power-law with index $ n $.
Although the index $ n_\mathrm s $ has been observed nearly scale-invariant at the Ly$ \alpha $F scale, we cannot assert that it also applies on small scales.
Hence, the footing where we stand in this section is that we regard the spectral index as a free parameter.
Then blue spectrum with $ n > 1 $ can produce PBHs on small scales.
Even if it is not the case, there is little reason to believe the absence of small-scale large density perturbation, which could be produced by some inflation mechanisms like double inflation.

Overproduced PBHs generally cause various problems in each epoch of the Universe, hence their abundance at each scale is required to be sufficiently small.
Among known constraints on PBH, the most stringent ones are set by cosmic rays assumed to be emitted via the Hawking-radiation process.
At each early epoch in the cosmic history, the PBH abundance is constrained so that energetic cosmic-rays from holes must not spoil the standard big-bang cosmology, for example, radiation entropy \cite{1978PThPh..59.1012M}, nucleosynthesis \cite{Kohri:1999ex}, 
and so on.
Also in the context of braneworld cosmology, interesting issues arise from cosmic rays;
the five-dimensional nature should be imprinted in the cosmic-ray spectra emitted from PBHs smaller than the anti--de Sitter curvature scale $ \ell $.
In this paper, we focus on a directly observable cosmic-ray in the present epoch, the extragalactic diffuse photon background.
The most dominant contribution to the spectrum comes from those holes currently evaporating.
Because of the red-shift dilution, contribution from the short-lived ones is suppressed, whereas low temperature of long-lived ones makes their emission power low.
The upper bound on $ \ell $ has been set around $ \sim 0.1~\text{mm} $ from gravitational experiments \cite{Hoyle:2000cv,Long:2002wn}.
As we will see, a lower border exists around $ \sim 10^{-11}~\text{m} $, below which PBHs whose lifetime is equal to the age of the Universe have larger radius than the five-dimensional scale.
In this case, five-dimensional effects on the observable cosmic-ray spectra are expected to diminish.

The strategy is summarized as follows.
(i) First we should determine of which size black holes take responsibility for the cosmic-ray spectra under consideration.
In the RS cosmology, the mass scale is affected by the size of the bulk curvature scale $ \ell $.
(ii) Then the abundance of such holes is constrained by the cosmic-ray observations.
(iii) From it we obtain constraints on the density perturbation using the mass spectrum formula.
To formulate cosmic-ray spectra, we describe Hawking evaporation of PBHs in the next subsection.

\subsection{Braneworld black hole and its evaporation}

In braneworld scenarios with some large compactification scale like Randall-Sundrum, a massive object localized on the brane within a typical scale should see the background space-time as an effectively flat five-dimensional Minkowski space \cite{Wiseman:2001xt,Kudoh:2003xz} (for possible deviations and specific exotic features, see \cite{Casadio:2002uv,Casadio:2001wh,Dadhich:2000am,Dadhich:2001ry,Bruni:2001fd,Govender:2001gq,Aliev:2005bi}).
Therefore a sufficiently small black hole is no longer ordinary 4D one, but is approximately described as 5D solution.
If there is no charge and is not rotating, the solution should be five-dimensional Schwarzschild.
The 5D Schwarzschild metric is given by
\begin{equation}
\mathrm ds^2
 = - \left[1 - \left(\frac{r_\mathrm S}{r}\right)^2\right] \mathrm dt^2
   + \left[1 - \left(\frac{r_\mathrm S}{r}\right)^2\right]^{-1} \mathrm dr^2
   + r^2 \mathrm d\Omega_3^2,
\end{equation}
where $ r_\mathrm S $ is the Schwarzschild radius and $ \mathrm d\Omega_n $ denotes the line element on unit $ n $-sphere.
The four-dimensional geometry seen by brane matter is obtained as its projection, i.e., after replacing $ \mathrm d\Omega_3 $ with $ \mathrm d\Omega_2 $.
In our framework the Schwarzschild radius in the five-dimensional spacetime is related to the black hole mass $ M_\mathrm{bh} $ and five-dimensional fundamental mass scale $ M_5 $ as \cite{Myers:1986un,Giddings:2001bu}
\begin{equation}
r_\mathrm{S}
 = \left(\frac{8}{3\pi}\frac{M_\mathrm{bh}}{M_5^3}\right)^{1/2}
 = \sqrt{\frac{8}{3\pi}}
   \left(\frac{\ell}{\ell_4}\right)^{1/2}
   \left(\frac{M_\mathrm{bh}}{M_4}\right)^{1/2} \ell_4,
\end{equation}
while in four dimensions $ r_\mathrm{S} = 2 (M_\mathrm{bh}/M_4) \ell_4 $.
The Hawking temperature of a Schwarzschild black hole is given in terms of its radius as $ T_\mathrm{H} = (D-3)/4\pi r_\mathrm{S} $, where $ D $ is the number of dimensions \cite{Myers:1986un}.
Thus in five dimensions
\begin{equation}
T_\mathrm{H}
 = \frac{1}{2\pi r_\mathrm{S}}
 = \sqrt{\frac{3}{32 \pi}}
   \left(\frac{\ell}{\ell_4}\right)^{-1/2}
   \left(\frac{M_\mathrm{bh}}{M_4}\right)^{-1/2} T_4,
\end{equation}
while in four dimensions $ T_\mathrm H = 1/(4\pi r_\mathrm S) = (1/8\pi) (M_\mathrm{bh}/M_4)^{-1} T_4 $.
For later use we define the effective Schwarzschild radius for massless fields \cite{Emparan:2000rs}
\begin{equation}
r_{\mathrm S,\mathrm{eff}}
 = \left(\frac{D-1}{2}\right)^{1/(D-3)} \left(\frac{D-1}{D-3}\right)^{1/2}
   r_\mathrm S.
\end{equation}
For $ D = 5 $ and $ D = 4 $, $ r_{\mathrm S,\mathrm{eff}} = 2 r_\mathrm S $ and $ r_{\mathrm S,\mathrm{eff}} = (3\sqrt{3}/2) r_\mathrm S $, respectively.
The surface area of the hole, $ A_D $, is
\begin{equation}
A_D
 = \Omega_{D-2} r_\mathrm S^{D-2},
\end{equation}
where $ \Omega_{D-2} $ is the area of the $ (D-1) $-dimensional unit sphere.
The effective area $ A_{D,\mathrm{eff}} $ is given by replacing the radius with $ r_{\mathrm S,\mathrm{eff}} $.

Next we describe the Hawking evaporation process of braneworld primordial black holes.
Recently numerical computations have been performed for emissions of matters into four dimensions \cite{Harris:2003eg,Kanti:2004nr} and that of graviton into bulk \cite{Cardoso:2005vb,Cardoso:2005mh,Park:2005vw,Creek:2006ia}, which are extensions of the four-dimensional result by Page \cite{Page:1976df}.
For the emission of brane matter, the calculation is done with the 4D metric obtained by projection, where the deviation from the 4D result comes through the form of metric function.
On the other hand, graviton emission is given using original 5D metric.
The number of dimensions of momentum space is denoted as $ d $, which we set $ 4 $ for brane matter or $ 5 $ for bulk graviton later.
The differential emission rate for general particle species, indicated by $ j $, is given by
\begin{equation}
\mathrm d\frac{\mathrm dN_j}{\mathrm dt}
 = g_j \frac{\sigma_j}{e^{E/T_\mathrm H} \pm 1}
   \frac{\mathrm dk^{d-1}}{(2 \pi)^{d-1}},
\label{eq:hr}
\end{equation}
where $ g_j $ is the number of internal degrees of freedom of the particle, $ E = (k^2 + m_j^2)^{1/2} $ is the energy, $ \sigma_j $ is the absorption cross section depending on hole's mass, particle's mass, energy, and spin.
The sign in the denominator expresses which statistics the particle obeys.
Numerical results showed that in the high-energy regime, i.e., when the geometric optics approximation would be applied, the absorption cross section is given by $ \sigma_j \simeq A_{d,\mathrm{eff}} \Omega_{d-3}/[(d-2) \Omega_{d-2}] $, but in the low-energy regime a certain amount of deviation exists \cite{Harris:2003eg,Cardoso:2005vb,Cardoso:2005mh,Park:2005vw,Creek:2006ia}.
The total mass-loss rate, Stefan-Boltzmann's law for black hole, becomes
\begin{equation}
\left(\frac{\mathrm dM_\mathrm{bh}}{\mathrm dt}\right)_\mathrm{evap}
 = -\sum_j \int_{m_j}^\infty
   E \frac{\mathrm d^2N_j}{\mathrm dE \mathrm dt} \mathrm dE
 = - \gamma_D(T_\mathrm H) \sigma_d A_{d,\mathrm{eff}} T_\mathrm H^d
 = - \frac{\epsilon(T_\mathrm H)}{r_\mathrm S^2},
\end{equation}
where $ \sigma_d $ is a $ d $-dimensional analog of Stefan-Boltzmann constant given with gamma function $ \Gamma(d) $ and zeta function $ \zeta(d) $ as
\begin{equation}
\sigma_d
 = \frac{\Omega_{d-3} \Gamma(d) \zeta(d)}{(d-2) (2 \pi)^{d-1}}
\end{equation}
and $ \gamma_D(T_\mathrm H) $ is the effective degree of freedom fixed after numerical result, which is absorbed into $ \epsilon $ for convenience
\begin{equation}
\epsilon(T_\mathrm H)
 = \gamma_D(T_\mathrm H) \Omega_{d-2} \sigma_d
   \left(\frac{d-1}{2}\right)^{(d-2)/(d-3)}
   \left(\frac{d-1}{d-3}\right)^{(d-2)/2}
   \left(\frac{d-3}{4\pi}\right)^d.
\end{equation}
To illuminate the effect of dimensionality, we rewrite it as
\begin{equation}
\epsilon
 = 1.118 \times 10^{-3} \sum_s g_{d,s}(T_\mathrm H) f_s \C_{D,d,s},
\end{equation}
where $ g_{d,s} $ is the effective number of dof for each spin $ s $ depending on the temperature, $ f_s $'s are those introduced in \cite{1991PhRvD..44..376M} expressing (relative) spin-dependent deviation in four dimensions from black-body emission, and $ \C_{D,d,s} $'s are the ratio of emission power relative to that in 4D.
We use Table~II--III of \cite{Harris:2003eg} for $ \C_{D,d=4,s \neq 2} $ to describe matter emission into brane and Table~I of \cite{Cardoso:2005mh} for $ \C_{D,d=5,s=2} $, bulk graviton emission\footnote{
The values for graviton in \cite{Cardoso:2005mh} includes polarization dof.
}.
Parameters we need are listed below
\begin{multline}
f_{s=1/2} =
\begin{cases}
0.147 & \text{uncharged} \\
0.142 & \text{with unit charge}
\end{cases},
f_{s=1} = 0.060,
f_{s=2} = 0.007, \\
\C_{D=5,d=4,s=1/2} = 14.2,
\C_{D=5,d=4,s=1} = 27.1,
\C_{D=5,d=5,s=2} = \frac{2}{5} \times 103.
\end{multline}
The numerical factor $ 1.118 \times 10^{-3} $ keeps the four-dimensional normalization of $ f_s $'s such that massless dof gives $ 2 \times f_{s=1} \times 1 + 6 \times f_{s=1/2} \times 1 = 1 $, which actually recovers $ (\mathrm dM_\mathrm{bh}/\mathrm dt)_\mathrm{evap} = - 5.34 \times 10^{25} (M_\mathrm{bh}/1~\text{g})^{-2}~\text{g}~\text{s}^{-1} $ \cite{1991PhRvD..44..376M}.
Since we are now in the braneworld, the degrees of freedom on the brane, namely the standard model particles, takes 3D phase space of momentum and black hole surface area is 2D, while that in the bulk, graviton, takes 4D phase space and the area is 3D.
Thus the mass-loss rate is formally given as
\begin{equation}
\left(\frac{\mathrm dM_\mathrm{bh}}{\mathrm dt}\right)_\mathrm{evap}
 = - \gamma_4 \sigma_4 A_{4,\mathrm{eff}} T_\mathrm H^4
   - \gamma_5 \sigma_5 A_{5,\mathrm{eff}} T_\mathrm H^5,
\end{equation}
where $ g_4 $ is calculated using the degree of freedom of the standard model matter, while $ g_5 $ is the degree of freedom of graviton polarization reflecting dimensionality as $ g_{s=2,d} = d(d-3)/2 $.
The above formula is apparently the same as in four dimensions except the added five-dimensional term for graviton dof.
However, the Schwarzschild radius does no longer obey four-dimensional relation for those holes much smaller than the compactification scale $ \ell $.
For a small enough black hole,
\begin{equation}
\left(\frac{\mathrm dM_\mathrm{bh}}{\mathrm dt}\right)_\mathrm{evap}
 = - \alpha_5
   \left(\frac{\ell}{\ell_4}\right)^{-1}
   \left(\frac{M_\mathrm{bh}}{M_4}\right)^{-1} \frac{M_4}{t_4},
\end{equation}
where the numerical factor is calculated according to the temperature of the hole as
\begin{equation}
\alpha_5
 \equiv \frac{3\pi}{8} \epsilon
 \approx
\begin{cases}
0.0227 & \gamma, \nu\text{'s}, \bar\nu\text{'s}, g \\
0.0333 & \gamma, \nu\text{'s}, \bar\nu\text{'s}, g, e, e^+
\end{cases}.
\end{equation}
Through a period in which $ \alpha_5 $ is regarded as constant, the solution is obtained as
\begin{equation}
M_\mathrm{bh}(t)
 = \left[
    \left(\frac{M_{\mathrm{bh},\mathrm i}}{M_4}\right)^2
    - 2 \alpha_5
      \left(\frac{\ell}{\ell_4}\right)^{-1}
      \left(\frac{t-t_\mathrm i}{t_4}\right)
 \right]^{1/2} M_4,
\end{equation}
where $ M_{\mathrm{bh},\mathrm i} = M_\mathrm{bh}(t_\mathrm i) $ is introduced as some initial condition.
The duration of this phase can be estimated as
\begin{equation}
\Delta t
 = \frac{1}{2 \alpha_5}
   \left(\frac{\ell}{\ell_4}\right)
   \left(\frac{M_{\mathrm{bh},\mathrm i}}{M_4}\right)^2 t_4.
\end{equation}
For comparison, we also show the 4D relations below
\begin{align}
\left(\frac{\mathrm dM_\mathrm{bh}}{\mathrm dt}\right)_\mathrm{evap}
 & = -\alpha_4 \left(\frac{M_\mathrm{bh}}{M_4}\right)^{-2}
     \frac{M_4}{t_4}, \\
\alpha_4
 & \equiv \frac{\epsilon}{4} 
 \approx
\begin{cases}
0.000283 & \gamma,\nu\text{'s},\bar\nu\text{'s}, g \\
0.000442 & \gamma,\nu\text{'s},\bar\nu\text{'s}, g, e, e^+
\end{cases}, \\
M_\mathrm{bh}(t)
 & = \left[
      \left(\frac{M_{\mathrm{bh},\mathrm i}}{M_4}\right)^3
      - 3 \alpha_4
        \left(\frac{t-t_\mathrm{i}}{t_4}\right)
     \right]^{1/3} M_4, \\
\Delta t
 & = \frac{1}{3 \alpha_4} \left(\frac{M_{\mathrm{bh},\mathrm i}}{M_4}\right)^3.
\end{align}

As we will discuss photons emitted from currently evaporating PBHs, our attention is concentrated on sufficiently large $ \ell $ cases in which there exist 5D PBHs with lifetimes comparable to the present age of the Universe.
For a typical 5D black hole with lifetime $ t_0 = 13.7~\text{Gyr} $, we can assume their temperature is low enough so that its emission channel is restricted to massless species and, at most, $ e/e^+ $.
Then its primordial mass $ M_{\mathrm{bh},\mathrm p}^* $ is calculated as
\begin{align}
M_{\mathrm{bh},\mathrm p}^*
 \approx \left[2 \alpha_5 \left(\frac{\ell}{\ell_4}\right)^{-1}
                          \left(\frac{t_0}{t_4}\right)\right]^{1/2} M_4
 \sim 5 \times 10^{9}
        \left(\frac{\ell}{0.1~\text{mm}}\right)^{-1/2}~\text{g}.
\label{eq:mass_evap}
\end{align}
Its corresponding temperature is
\begin{equation}
T_\mathrm H^*
 \sim 50 \left(\frac{\ell}{0.1~\text{mm}}\right)^{-1/4}~\text{keV},
\end{equation}
from which we confirm the consistency of low-energy approximation.
The dominant contribution to the currently observable photon spectrum comes from this $ M_{\mathrm{bh},\mathrm p}^* $ holes.
The condition for the five dimensionality of $ M_{\mathrm{bh},\mathrm p}^* $ PBH is set as
\begin{equation}
r_\mathrm S(M_{\mathrm{bh},\mathrm p}^*)
 \sim 10^{-9} \left(\frac{\ell}{0.1~\text{mm}}\right)^{-1/4}~\text{mm}
 \lesssim \ell.
\end{equation}
This gives $ \ell \gtrsim 10^{-11}~\text{m} $.
On the other hand, such 5D PBHs can be formed only when the reheating temperature is sufficiently high such that at least $ f M_{\mathrm h,\mathrm{rh}} < M_{\mathrm{bh},\mathrm p}^* $.
This condition reads
\begin{equation}
T_\mathrm{rh}
 \gtrsim 10^5 f^{1/8} \left(\frac{\ell}{0.1~\text{mm}}\right)^{-5/16}~\text{GeV},
\end{equation}
which is, for large $ \ell $, stronger than Eq.~(\ref{eq:T_rhmin}).
The presence of accretion, $ F > 0 $, makes this lower limit stringent.
$ F = 1 $ with $ \ell = 0.1~\text{mm} $ and $ f = 1 $ gives $ T_\mathrm{rh} \gtrsim 10^6~\text{GeV} $.
In the 4D case, we get $ T_\mathrm{rh} \gtrsim 10^8 f^{1/2}~\text{GeV} $ for the same argument.

\subsection{Constraints from the extragalactic diffuse photon background}

It is not the purpose of this paper to show constraints on perturbation spectrum for the whole region of the parameter space.
Rather, we discuss the most remarkable case, namely $ \ell = 0.1~\text{mm} $, and see how the existence of the extra dimension modifies constraints on the perturbation from four dimensions.
In addition, two distinguishable cases for accretion efficiency, $ F = 1 $ and $ F = 0 $, will be examined.
We give constraints on the set of the spectral index $ n $ and the reheating temperature $ T_\mathrm{rh} $.
We calculate the photon surface brightness for each set and determine if the set is allowed or not by the criterion that the whole spectrum is below the observed data or not.
The photon background data used here are taken from \cite{1999ApJ...520..124G,2000AIPC..510..467W,Strong:2004ry,Strong:2004de} (we also show the gamma-ray result in \cite{Sreekumar:1997un}).

Let us formulate the diffuse photon spectrum from PBHs.
At some fixed time $ t $, the photon emission power is given as
\begin{equation}
\frac{\mathrm dU(E)}{\mathrm dt}
 \approx \int \mathrm dM_{\mathrm{bh},\mathrm p}
         (1+z)^3
         \frac{\mathrm dn_\mathrm{bh}}{\mathrm dM_{\mathrm{bh},\mathrm p}} E^2
         \frac{\mathrm d^2N_\gamma}{\mathrm dE \mathrm dt},
\end{equation}
where $ (1+z)^3 $ expresses the decreasing rate of number density.
For the emission rate per hole, we take $ g_\gamma = 2 $ and $ d = 4 $ for Eq.~(\ref{eq:hr}) and use geometric optics approximation
\begin{equation}
\frac{\mathrm d^2N_\gamma}{\mathrm dE \mathrm dt}
 \approx \frac{\pi r_{\mathrm S,\mathrm{eff}}^2}{e^{E/T_\mathrm H} - 1}
   \frac{E^2}{\pi^2}.
\end{equation}
Here we ignore secondary photons.
Photons suffer red-shift and the energy that we will observe is $ E_0 = E/(1+z) $.
Energy density $ U $ decreases as $ (1+z)^4 $ to give its present density $ U_0 $
\begin{equation}
U_0(E_0)
 = \int_{t_\mathrm{dec}}^{t_0} \mathrm dt (1+z)^{-4}
   \frac{\mathrm dU(E)}{\mathrm dt}
 = \int \mathrm dM_{\mathrm{bh},\mathrm p}
   \frac{\mathrm dn_\mathrm{bh}}{\mathrm dM_{\mathrm{bh},\mathrm p}}
   \int_{t_\mathrm{dec}}^{t_\mathrm{evap}} \mathrm dt \frac{E^2}{1+z}
   \frac{\mathrm d^2N_\gamma}{\mathrm dE \mathrm dt}.
\end{equation}
An observable quantity which we use is surface brightness given as
\begin{equation}
I(E_0)
 = \frac{c}{4\pi} \frac{U_0(E_0)}{E_0}
   ~ \text{keV}~\text{cm}^{-2}~\text{sr}^{-1}~\text{s}^{-1}~\text{keV}^{-1}.
\end{equation}

\begin{figure}[ht]
\begin{center}
\includegraphics{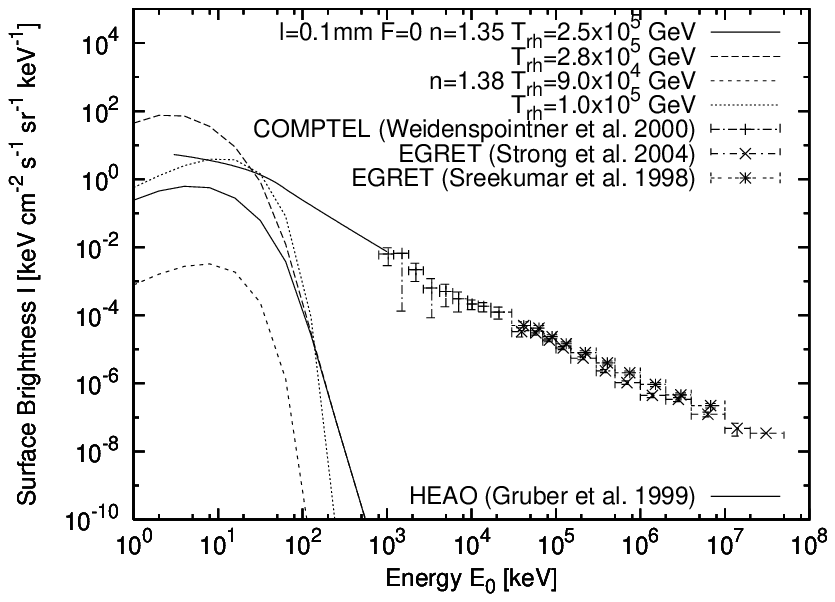}
\includegraphics{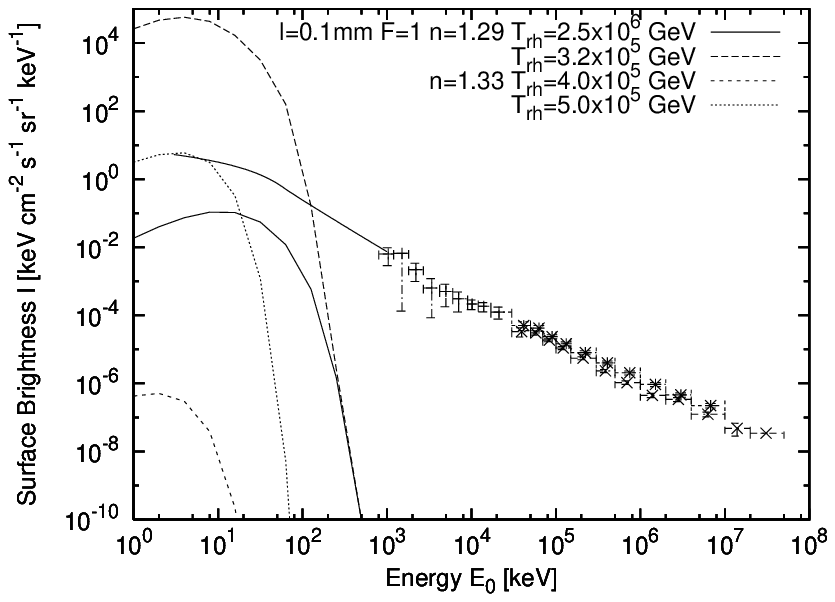}
\includegraphics{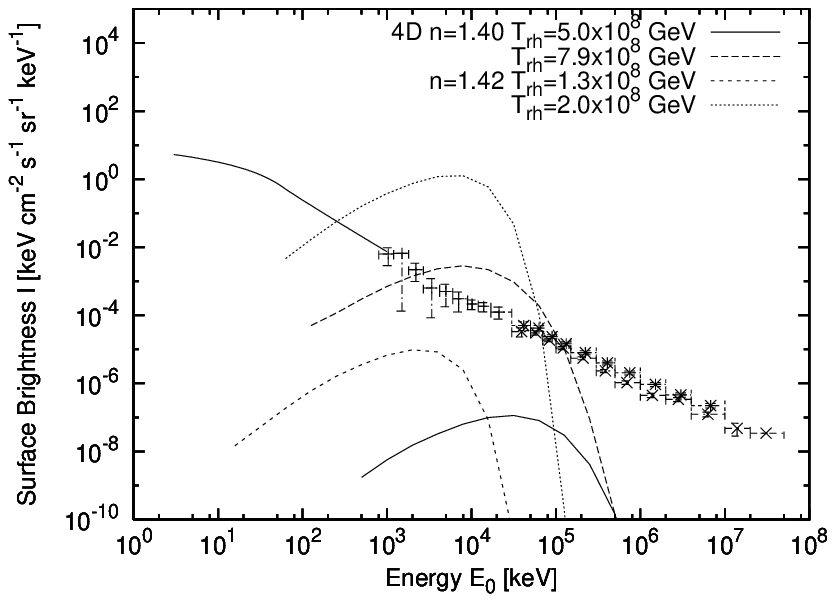}
\caption{
Illustration of constraints on the photon spectrum.
Each panel corresponds to a combination of $ \ell $ and $ F $.
For each $ n $, one of the spectra for lower $ T_\mathrm{rh}$ is allowed and the other for higher $ T_\mathrm{rh} $ is rejected.
}
\end{center}
\label{fig:illust}
\end{figure}

We observe that there are two possibilities one should take care:
exponential spectrum or power-law spectrum around $ M_\mathrm{bh}^* $ according to the spectral index $ n \gtrsim n^* $ or $ n \lesssim n^* $, respectively, where $ n^* $ is given by equating $ \delta_{\mathrm h,\mathrm{min}} \sim \sigma_\mathrm h(M_\mathrm{bh}^*) $ as
\begin{equation}
n^* - 1
 = \frac{\log[\delta_{\mathrm h,\mathrm{min}}/(3/5) \sigma_0]}
        {\log\left[(M_{\mathrm{bh},\mathrm i}^*/f)^{-3/16} M_\mathrm c^{-1/16}
                    M_{\mathrm h,0}^{1/4}\right]}
 \simeq \frac{4.0}{8.4 + 3.0y + (0.014+0.12y) \log(\ell/0.1~\text{mm})
                       + (0.08+0.08y) \log f},
\label{eq:nul}
\end{equation}
where $ y \equiv F/(\pi-F) $.
In the case of $ \ell = 0.1~\text{mm} $ and $ f = 1 $, this gives $ n^* \approx 1.5 $ for $ F = 0 $ and $ n^* \approx 1.4 $ for $ F = 1 $.
First we examine the possibility of $ n > n^* $.
In general such a large $ n $ causes excess above the observed photon background, so it appears that $ n > n^* $ is rejected.
However, we should remember that the lightest mass of existing PBHs, $ M_{\mathrm{bh},\mathrm{min}} = f M_{\mathrm h,\mathrm{rh}} $, is determined by reheating temperature.
If $ T_\mathrm{rh} $ drops below $ \sim 100 f^{1/8} (\ell/0.1~\text{mm})^{-5/16}~\text{TeV} $, then the mass of the lightest PBH goes above $ M_\mathrm{bh}^* $ (see Eqs.~(\ref{eq:mass_rh}) and (\ref{eq:mass_evap})) and the constraint from the photon background makes no sense.
On the other hand, $ n \ll n^* $ case is not constrained any more since high reheating temperature only increases lighter PBHs than $ M_{\mathrm{bh},\mathrm p}^* $ which virtually does not contribute to the photon background.
The problem reduces to a question how far the upper limit of $ n $ can approach $ n^* $.
Some illustrative examples of the photon spectrum are shown in Fig.~\ref{fig:illust}.
The case with the large extra dimension, $ \ell = 0.1~\text{mm} $, with two possibilities of the accretion efficiency, $ F = 1 $ and $ F = 0 $, are considered.
For each set of $ (\ell,F) $, four photon spectra for different perturbation spectral index $ n $ are shown;
two are allowed by either smaller spectral index or lower reheating temperature, while the other two are rejected for either larger index or higher temperature.
From these figures, we observe that the constraint on the spectral index is the strongest for $ (\ell=0.1~\text{mm},F=1) $.
This is because the mass spectrum is the hardest for this case so that the holes with mass around $ M_\mathrm{bh}^* $ are potentially most abundant.

Fig.~\ref{fig:illust} shows that every spectrum in the 5D cases makes a peak in the energy region of the High Energy Astrophysics Observatory (HEAO) \cite{1999ApJ...520..124G}.
One notices that the spectra are significantly red-shifted compared to those presented in our previous work \cite{Sendouda:2003dc}, which were rather similar to black body emission with temperature $ T_\mathrm H^* $.
This reflects two facts;
one is that the spectra in \cite{Sendouda:2003dc} were drawn assuming scale-invariant power spectrum, whereas it is turned to be not allowed in the present setup from the above arguments.
An exponentially decaying mass spectrum leads to a relative reduction of the population of heavier PBHs than $ M_\mathrm{bh}^* $.
The other comes from lowered reheating temperature, which reduces the number of lighter PBHs than $ M_\mathrm{bh}^* $.
Both the two facts indicate that the contribution to photon spectra from holes other than $ M_{\mathrm{bh},\mathrm p}^* $ is enhanced, hence the superposed photon spectra are seen to be relatively red-shifted.
Note that even if the perturbation index $ n $ is more blue, say $ n \sim 2 $, lower reheating temperature could save the situation by virtue of this ``reddening'' effect.
Of course this is only the consequence of the assumptions in our method, namely, the single power-law density perturbation spectrum with the large-scale WMAP+SDSS-gal+Ly$ \alpha $F normalization.

\begin{figure}[ht]
\begin{center}
\includegraphics{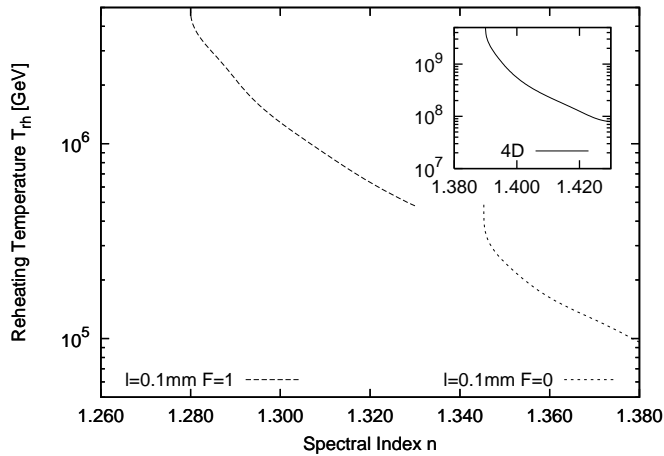}
\caption{
Allowed regions of $ (n,T_\mathrm{rh}) $ for each combination of $ (\ell,F) $.
Left-hand side is allowed and the opposite is rejected.
The 4D result is also shown in the inner panel.
}
\end{center}
\label{fig:result}
\end{figure}

At last, the allowed region for the set $ (n,T_\mathrm{rh}) $ is given in Fig.~\ref{fig:result}.
As was mentioned previously, any $ n $ could be allowed no matter how it is large if only the reheating temperature is low enough.
However, we here focus on the most meaningful region, that is, around the possible upper bound on the index $ n $ in the case of marginally high $ T_\mathrm{rh} $.
Approximately it is sufficient to require either conditions on $ T_\mathrm{rh} $ or $ n $ is satisfied.
The constraints on the braneworld inflation for $ \ell = 0.1~\text{mm} $ is summarized in Tab.~\ref{tab:result}.
\begin{table}[htb]
\begin{center}
\caption{
Requirements for the reheating temperature $ T_\mathrm{rh} $ and spectral index $ n $.
}
\label{tab:result}
\begin{tabular}{c|cccc}
Dimensionality/Accretion
 &
 & Threshold
 &
 & To realize larger $ n $: \\
\hline
$ \ell=0.1~\text{mm} $/$ F = 0 $
 &
 & $ n \approx 1.35 $
 &
 & $ T_\mathrm{rh} \lesssim 10^5~\text{GeV} $ \\
$ \ell=0.1~\text{mm} $/$ F = 1 $
 &
 & $ n \approx 1.28 $
 &
 & $ T_\mathrm{rh} \lesssim 10^6~\text{GeV} $ \\
4D
 &
 & $ n \approx 1.39 $
 &
 & $ T_\mathrm{rh} \lesssim 10^8~\text{GeV} $
\end{tabular}
\end{center}
\end{table}
Although only the case of $ \ell = 0.1~\text{mm} $ has been considered, we can anticipate what general $ \ell $ results in.
Considering Eq.~(\ref{eq:nul}), we deduce that both smaller $ \ell $ and smaller $ F $ will weaken the constraint.

We also obtained an updated constraint on four-dimensional PBHs.
Since we used the observational data of the small-scale density perturbation obtained by WMAP, SDSS, and Ly$ \alpha $ Forest, the constraint on the spectral index $ n $ is relaxed compared to the result \cite{Kim:1999iv} which used the COBE data normalized at the present horizon-scale.
On the other hand, the border of the reheating temperature is unchanged.

\section{\label{sec:conclusion}
Discussion and conclusion
}

In this paper, we obtained the primordial black hole mass spectrum produced from inflationary density perturbation in the Randall-Sundrum braneworld.
The result is Eqs.~(\ref{eq:ms1}) and (\ref{eq:ms2}).
Fig.~\ref{fig:ms} showed typical shapes of the mass spectrum.
Apart from the normalization, the mass spectrum contains two scales.
One corresponds to the scale of the transition from the $ \rho $-square to the ordinary radiation-dominated era.
Below this scale the mass spectrum is deformed.
In the case of no accretion, $ F = 0 $, the index is softened to $ -17/8 $ from the four-dimensional value $ -5/2 $.
On the other hand, in the presence of accretion, $ F = 1 $, the index is hardened to $ -2.65 $.
The other scale in the mass spectrum is of the transition from power-law to exponential decreasing specified by $ \delta_{\mathrm h,\mathrm{min}} \sim \sigma_\mathrm h(M_\mathrm{th}) $.

Using above results, we calculated the diffuse photon spectrum emitted from PBHs to constrain the perturbation spectrum $ n $ and/or the reheating temperature $ T_\mathrm{rh} $.
We considered a power-law power spectrum of curvature perturbation $ \P_{\R_\mathrm c} = A (k/k_0)^{n-1} $ which is normalized at the Lyman-$ \alpha $ forest scale $ k_0 = 0.9 h~\text{Mpc}^{-1} $ to be $ A = 2.28 \times 10^{-9} $.
Recent numerical results on the Hawking radiation in higher-dimensions \cite{Harris:2003eg,Kanti:2004nr,Cardoso:2005vb,Cardoso:2005mh,Cornell:2005ux,Park:2005vw,Creek:2006ia} were incorporated.
The observed photon background placed an upper limit on the photons from PBHs.
Since the contribution to the photon spectrum is dominated by currently evaporating holes, the constraint was set so that the amount of such critical holes be sufficiently suppressed.
Fig.~\ref{fig:illust} showed illustrations of spectra.
By virtue of the rapidly varying form of the mass spectrum, the result presented in Fig.~\ref{fig:result} and Tab.~\ref{tab:result} were summarized simply:
The requirement is either the spectral index $ n $ is smaller than some critical value, or the reheating temperature $ T_\mathrm{rh} $ is lower than some critical value.
This upper limit for $ n $ could be estimated by $ n^* $ in Eq.~(\ref{eq:nul}), above which the mass spectrum around $ M_{\mathrm{bh},\mathrm p}^* $ becomes power-law, while the threshold for $ T_\mathrm{rh} $ was given from the condition $ f M_\mathrm h(T_\mathrm{rh}) = M_{\mathrm{bh},\mathrm p}^* $, below which $ M_{\mathrm{bh},\mathrm p} $ holes cannot be produced.
It is interesting to note that the values of $ T_\mathrm{rh} $ for five dimension cases in Tab.~\ref{tab:result} is comparable to those values obtained for unstable gravitino decay \cite{Kawasaki:2004yh,Kawasaki:2004qu,Kohri:2005wn}, which are not connected with the dimensionality.

It was helpful that each formula has been given a form which manifestly shows the recovery of the standard four-dimensional version (``$ \rho^1 $'' part) and the deviation caused by the higher dimension (``$ \rho^2 $'' part).
Taking the recent observational data into account, we also have successfully updated the four-dimensional result on this subject obtained in the literature \cite{Kim:1999iv}.

In the formulation, we owed a simplification with respect to the generation and evolution of super-horizon curvature fluctuation in the braneworld cosmology.
There are still difficulties to handle the perturbation on the brane considering the bulk geometry, but some intuition can be made from the results of tensor perturbations \cite{Kobayashi:2003cn,Hiramatsu:2003iz,Hiramatsu:2004aa,Kobayashi:2005jx,Kobayashi:2005dd,Kinoshita:2005nx,Hiramatsu:2006bd,Kobayashi:2006pe,Seahra:2006tm}.
Gravitational waves produced in the inflationary regime or matter gravitational collapse can escape into the bulk, so the energy-loss into the bulk can also takes an important role in the formation process of primordial black holes.
In this sense, omitting the perturbation of bulk geometry might lead us to an over estimate of the PBH production, and the constraint obtained in this paper may be weaken after the effects will be appropriately taken into account.

The inflation mechanism itself in the braneworld scenario is rather unveiled.
There are some suggestion for the brane inflation and generation of the density perturbation on the brane \cite{Maartens:1999hf,Koyama:2003yz}, and we have had some constraints on the spectrum in the literature \cite{Liddle:2003gw,Tsujikawa:2003zd}.
PBH offers an alternative way to investigate brane inflations.

\begin{acknowledgments}
Y.S. and S.N. are pleased to acknowledge helpful discussions with Mamoru Shimizu.
The authors would also like to thank Hee Il Kim, Shunichiro Kinoshita, Hidetaka Sonoda, Jun'ichi Yokoyama, Masahide Yamaguchi, Akihito Shirata, and Takashi Hiramatsu for useful comments.
This work was supported in part through Grant-in-Aid for Scientific Research (S) No.~14102004, Grant-in-Aid for Scientific Research on Priority Areas No.~14079202, and Grant-in-Aid for Scientific Research No.~16740134 by Japan Ministry of Education, Culture, Sports, Science and Technology.
Y.S. is supported by Japan Society for the Promotion of Science.
\end{acknowledgments}

\bibliographystyle{hunsrt}
\bibliography{pbhms}

\begin{thebibliography}{100}

\bibitem{1967SvA....10..602Z}
Y.~B. {Zel'Dovich} and I.~D. {Novikov}.
\newblock {The Hypothesis of Cores Retarded during Expansion and the Hot
  Cosmological Model}.
\newblock {\em Soviet Astronomy}, 10:602--+, February 1967.

\bibitem{Hawking:1971ei}
Stephen Hawking.
\newblock Gravitationally collapsed objects of very low mass.
\newblock {\em Mon. Not. Roy. Astron. Soc.}, 152:75, 1971.

\bibitem{Carr:1974nx}
B.~J. Carr and S.~W. Hawking.
\newblock Black holes in the early universe.
\newblock {\em Mon. Not. Roy. Astron. Soc.}, 168:399--415, 1974.

\bibitem{Carr:1975qj}
Bernard~J. Carr.
\newblock The primordial black hole mass spectrum.
\newblock {\em Astrophys. J.}, 201:1--19, 1975.

\bibitem{Hawking:1974rv}
S.~W. Hawking.
\newblock Black hole explosions.
\newblock {\em Nature}, 248:30--31, 1974.

\bibitem{Hawking:1974sw}
S.~W. Hawking.
\newblock Particle creation by black holes.
\newblock {\em Commun. Math. Phys.}, 43:199--220, 1975.

\bibitem{Page:1976df}
Don~N. Page.
\newblock Particle emission rates from a black hole: Massless particles from an
  uncharged, nonrotating hole.
\newblock {\em Phys. Rev.}, D13:198--206, 1976.

\bibitem{Page:1976wx}
D.~N. Page and S.~W. Hawking.
\newblock Gamma rays from primordial black holes.
\newblock {\em Astrophys. J.}, 206:1--7, 1976.

\bibitem{1990PhRvD..41.3052M}
J.~H. {MacGibbon} and B.~R. {Webber}.
\newblock {Quark- and gluon-jet emission from primordial black holes: The
  instantaneous spectra}.
\newblock {\em \prd}, 41:3052--3079, May 1990.

\bibitem{1991PhRvD..44..376M}
J.~H. {MacGibbon}.
\newblock {Quark- and gluon-jet emission from primordial black holes. II. The
  emission over the black-hole lifetime}.
\newblock {\em \prd}, 44:376--392, July 1991.

\bibitem{1991ApJ...371..447M}
J.~H. {MacGibbon} and B.~J. {Carr}.
\newblock {Cosmic rays from primordial black holes}.
\newblock {\em \apj}, 371:447--469, April 1991.

\bibitem{1976ApJ...206....8C}
B.~J. {Carr}.
\newblock {Some cosmological consequences of primordial black-hole
  evaporations}.
\newblock {\em \apj}, 206:8--25, May 1976.

\bibitem{1987Natur.329..308M}
J.~H. {MacGibbon}.
\newblock {Can Planck-mass relics of evaporating black holes close the
  universe?}
\newblock {\em \nat}, 329:308--+, September 1987.

\bibitem{Sato:1980yn}
K.~Sato.
\newblock First order phase transition of a vacuum and expansion of the
  universe.
\newblock {\em Mon. Not. Roy. Astron. Soc.}, 195:467--479, 1981.

\bibitem{Guth:1980zm}
Alan~H. Guth.
\newblock The inflationary universe: A possible solution to the horizon and
  flatness problems.
\newblock {\em Phys. Rev.}, D23:347--356, 1981.

\bibitem{Carr:1994ar}
B.~J. Carr, J.~H. Gilbert, and James~E. Lidsey.
\newblock Black hole relics and inflation: Limits on blue perturbation spectra.
\newblock {\em Phys. Rev.}, D50:4853--4867, 1994, astro-ph/9405027.

\bibitem{Green:1997sz}
Anne~M. Green and Andrew~R. Liddle.
\newblock Constraints on the density perturbation spectrum from primordial
  black holes.
\newblock {\em Phys. Rev.}, D56:6166--6174, 1997, astro-ph/9704251.

\bibitem{Kim:1996hr}
Hee~Il Kim and Chul~H. Lee.
\newblock Constraints on the spectral index from primordial black holes.
\newblock {\em Phys. Rev.}, D54:6001--6007, 1996.

\bibitem{Kim:1999iv}
Hee~Il Kim, Chul~H. Lee, and Jane~H. MacGibbon.
\newblock Diffuse gamma-ray background and primordial black hole constraints on
  the spectral index of density fluctuations.
\newblock {\em Phys. Rev.}, D59:063004, 1999, astro-ph/9901030.

\bibitem{Bringmann:2001yp}
Torsten Bringmann, Claus Kiefer, and David Polarski.
\newblock Primordial black holes from inflationary models with and without
  broken scale invariance.
\newblock {\em Phys. Rev.}, D65:024008, 2002, astro-ph/0109404.

\bibitem{Blais:2002gw}
David Blais, Torsten Bringmann, Claus Kiefer, and David Polarski.
\newblock Accurate results for primordial black holes from spectra with a
  distinguished scale.
\newblock {\em Phys. Rev.}, D67:024024, 2003, astro-ph/0206262.

\bibitem{Rubin:2000dq}
S.~G. Rubin, M.~Yu. Khlopov, and A.~S. Sakharov.
\newblock Primordial black holes from non-equilibrium second order phase
  transition.
\newblock {\em Grav. Cosmol.}, S6:51--58, 2000, hep-ph/0005271.

\bibitem{Rubin:2001yw}
Sergei~G. Rubin, Alexander~S. Sakharov, and Maxim~Yu. Khlopov.
\newblock The formation of primary galactic nuclei during phase transitions in
  the early universe.
\newblock {\em J. Exp. Theor. Phys.}, 91:921--929, 2001, hep-ph/0106187.

\bibitem{Barrau:2002ru}
A.~Barrau, D.~Blais, G.~Boudoul, and D.~Polarski.
\newblock Galactic cosmic rays from pbhs and primordial spectra with a scale.
\newblock {\em Phys. Lett.}, B551:218--225, 2003, astro-ph/0210149.

\bibitem{Bugaev:2002yt}
E.~V. Bugaev and K.~V. Konishchev.
\newblock Cosmological constraints from evaporations of primordial black holes.
\newblock {\em Phys. Rev.}, D66:084004, 2002, astro-ph/0206082.

\bibitem{Randall:1999ee}
Lisa Randall and Raman Sundrum.
\newblock A large mass hierarchy from a small extra dimension.
\newblock {\em Phys. Rev. Lett.}, 83:3370--3373, 1999, hep-ph/9905221.

\bibitem{Randall:1999vf}
Lisa Randall and Raman Sundrum.
\newblock An alternative to compactification.
\newblock {\em Phys. Rev. Lett.}, 83:4690--4693, 1999, hep-th/9906064.

\bibitem{Cline:1999ts}
James~M. Cline, Christophe Grojean, and Geraldine Servant.
\newblock Cosmological expansion in the presence of extra dimensions.
\newblock {\em Phys. Rev. Lett.}, 83:4245, 1999, hep-ph/9906523.

\bibitem{Csaki:1999jh}
Csaba Csaki, Michael Graesser, Christopher~F. Kolda, and John Terning.
\newblock Cosmology of one extra dimension with localized gravity.
\newblock {\em Phys. Lett.}, B462:34--40, 1999, hep-ph/9906513.

\bibitem{Binetruy:1999ut}
Pierre Binetruy, Cedric Deffayet, and David Langlois.
\newblock Non-conventional cosmology from a brane-universe.
\newblock {\em Nucl. Phys.}, B565:269--287, 2000, hep-th/9905012.

\bibitem{Binetruy:1999hy}
Pierre Binetruy, Cedric Deffayet, Ulrich Ellwanger, and David Langlois.
\newblock Brane cosmological evolution in a bulk with cosmological constant.
\newblock {\em Phys. Lett.}, B477:285--291, 2000, hep-th/9910219.

\bibitem{Maartens:2003tw}
Roy Maartens.
\newblock Brane-world gravity.
\newblock {\em Living Rev. Rel.}, 7:7, 2004, gr-qc/0312059.

\bibitem{Guedens:2002km}
Raf Guedens, Dominic Clancy, and Andrew~R Liddle.
\newblock Primordial black holes in braneworld cosmologies. i: Formation,
  cosmological evolution and evaporation.
\newblock {\em Phys. Rev.}, D66:043513, 2002, astro-ph/0205149.

\bibitem{Guedens:2002sd}
Raf Guedens, Dominic Clancy, and Andrew~R. Liddle.
\newblock Primordial black holes in braneworld cosmologies: Accretion after
  formation.
\newblock {\em Phys. Rev.}, D66:083509, 2002, astro-ph/0208299.

\bibitem{Majumdar:2002mr}
A.~S. Majumdar.
\newblock Domination of black hole accretion in brane cosmology.
\newblock {\em Phys. Rev. Lett.}, 90:031303, 2003, astro-ph/0208048.

\bibitem{Clancy:2003zd}
Dominic Clancy, Raf Guedens, and Andrew~R. Liddle.
\newblock Primordial black holes in braneworld cosmologies: Astrophysical
  constraints.
\newblock {\em Phys. Rev.}, D68:023507, 2003, astro-ph/0301568.

\bibitem{Sendouda:2003dc}
Yuuiti Sendouda, Shigehiro Nagataki, and Katsuhiko Sato.
\newblock Constraints on the mass spectrum of primordial black holes and
  braneworld parameters from the high-energy diffuse photon background.
\newblock {\em Phys. Rev.}, D68:103510, 2003, astro-ph/0309170.

\bibitem{Sendouda:2004hz}
Yuuiti Sendouda, Kazunori Kohri, Shigehiro Nagataki, and Katsuhiko Sato.
\newblock Sub-gev galactic cosmic-ray antiprotons from pbhs in the
  randall-sundrum braneworld.
\newblock {\em Phys. Rev.}, D71:063512, 2005, astro-ph/0408369.

\bibitem{Maartens:1999hf}
Roy Maartens, David Wands, Bruce~A. Bassett, and Imogen Heard.
\newblock Chaotic inflation on the brane.
\newblock {\em Phys. Rev.}, D62:041301, 2000, hep-ph/9912464.

\bibitem{Himemoto:2000nd}
Yoshiaki Himemoto and Misao Sasaki.
\newblock Brane-world inflation without inflaton on the brane.
\newblock {\em Phys. Rev.}, D63:044015, 2001, gr-qc/0010035.

\bibitem{Koyama:2003yz}
Kazuya Koyama and Keitaro Takahashi.
\newblock Primordial fluctuations in bulk inflaton model.
\newblock {\em Phys. Rev.}, D67:103503, 2003, hep-th/0301165.

\bibitem{Bennett:2003bz}
C.~L. Bennett et~al.
\newblock First year wilkinson microwave anisotropy probe (wmap) observations:
  Preliminary maps and basic results.
\newblock {\em Astrophys. J. Suppl.}, 148:1, 2003, astro-ph/0302207.

\bibitem{Spergel:2003cb}
D.~N. Spergel et~al.
\newblock First year wilkinson microwave anisotropy probe (wmap) observations:
  Determination of cosmological parameters.
\newblock {\em Astrophys. J. Suppl.}, 148:175, 2003, astro-ph/0302209.

\bibitem{Seljak:2004xh}
Uros Seljak et~al.
\newblock Cosmological parameter analysis including sdss ly-alpha forest and
  galaxy bias: Constraints on the primordial spectrum of fluctuations, neutrino
  mass, and dark energy.
\newblock {\em Phys. Rev.}, D71:103515, 2005, astro-ph/0407372.

\bibitem{1974ApJ...187..425P}
W.~H. {Press} and P.~{Schechter}.
\newblock {Formation of Galaxies and Clusters of Galaxies by Self-Similar
  Gravitational Condensation}.
\newblock {\em \apj}, 187:425--438, February 1974.

\bibitem{Wands:2000dp}
David Wands, Karim~A. Malik, David~H. Lyth, and Andrew~R. Liddle.
\newblock A new approach to the evolution of cosmological perturbations on
  large scales.
\newblock {\em Phys. Rev.}, D62:043527, 2000, astro-ph/0003278.

\bibitem{Liddle:2000cg}
A.~R. Liddle and D.~H. Lyth.
\newblock {\em Cosmological inflation and large-scale structure}.
\newblock Cambridge, UK: University Press, 2000.
\newblock 400 p.

\bibitem{Koyama:2000cc}
Kazuya Koyama and Jiro Soda.
\newblock Evolution of cosmological perturbations in the brane world.
\newblock {\em Phys. Rev.}, D62:123502, 2000, hep-th/0005239.

\bibitem{Langlois:2000iu}
David Langlois, Roy Maartens, Misao Sasaki, and David Wands.
\newblock Large-scale cosmological perturbations on the brane.
\newblock {\em Phys. Rev.}, D63:084009, 2001, hep-th/0012044.

\bibitem{Koyama:2001ct}
Kazuya Koyama and Jiro Soda.
\newblock Bulk gravitational field and cosmological perturbations on the brane.
\newblock {\em Phys. Rev.}, D65:023514, 2002, hep-th/0108003.

\bibitem{Koyama:2004ap}
Kazuya Koyama, David Langlois, Roy Maartens, and David Wands.
\newblock Scalar perturbations from brane-world inflation.
\newblock {\em JCAP}, 0411:002, 2004, hep-th/0408222.

\bibitem{Yoshiguchi:2004nm}
Hiroyuki Yoshiguchi and Kazuya Koyama.
\newblock Quantization of scalar perturbations in brane-world inflation.
\newblock {\em Phys. Rev.}, D71:043519, 2005, hep-th/0411056.

\bibitem{Koyama:2005ek}
Kazuya Koyama, Shuntaro Mizuno, and David Wands.
\newblock Slow-roll corrections to inflaton fluctuations on a brane.
\newblock {\em JCAP}, 0508:009, 2005, hep-th/0506102.

\bibitem{Niemeyer:1999ak}
J.~C. Niemeyer and K.~Jedamzik.
\newblock Dynamics of primordial black hole formation.
\newblock {\em Phys. Rev.}, D59:124013, 1999, astro-ph/9901292.

\bibitem{Niemeyer:1997mt}
J.~C. Niemeyer and K.~Jedamzik.
\newblock Near-critical gravitational collapse and the initial mass function of
  primordial black holes.
\newblock {\em Phys. Rev. Lett.}, 80:5481--5484, 1998, astro-ph/9709072.

\bibitem{Yokoyama:1998xd}
Jun'ichi Yokoyama.
\newblock Cosmological constraints on primordial black holes produced in the
  near-critical gravitational collapse.
\newblock {\em Phys. Rev.}, D58:107502, 1998, gr-qc/9804041.

\bibitem{Tikhomirov:2005bt}
Victor~V. Tikhomirov and Y.~A. Tsalkou.
\newblock How particle collisions increase the rate of accretion from
  cosmological background onto primordial black holes in braneworld cosmology.
\newblock {\em Phys. Rev.}, D72:121301, 2005, astro-ph/0510212.

\bibitem{Ichiki:2002eh}
K.~Ichiki, M.~Yahiro, T.~Kajino, M.~Orito, and G.~J. Mathews.
\newblock Observational constraints on dark radiation in brane cosmology.
\newblock {\em Phys. Rev.}, D66:043521, 2002, astro-ph/0203272.

\bibitem{Kodama:1985bj}
Hideo Kodama and Misao Sasaki.
\newblock Cosmological perturbation theory.
\newblock {\em Prog. Theor. Phys. Suppl.}, 78:1--166, 1984.

\bibitem{Kawasaki:2004jd}
Masahiro Kawasaki.
\newblock Condition for primordial black hole formation in randall- sundrum
  cosmology.
\newblock {\em Phys. Lett.}, B591:203--207, 2004, astro-ph/0403668.

\bibitem{1978PThPh..59.1012M}
S.~{Miyama} and K.~{Sato}.
\newblock {An Upper Bound on the Number Density of Primordial Black Holes from
  the Big Bang Nucleosynthesis}.
\newblock {\em Progress of Theoretical Physics}, 59:1012--1013, March 1978.

\bibitem{Kohri:1999ex}
K.~Kohri and Jun'ichi Yokoyama.
\newblock Primordial black holes and primordial nucleosynthesis. i: Effects of
  hadron injection from low mass holes.
\newblock {\em Phys. Rev.}, D61:023501, 2000, astro-ph/9908160.

\bibitem{Hoyle:2000cv}
C.~D. Hoyle et~al.
\newblock Sub-millimeter tests of the gravitational inverse-square law: A
  search for 'large' extra dimensions.
\newblock {\em Phys. Rev. Lett.}, 86:1418--1421, 2001, hep-ph/0011014.

\bibitem{Long:2002wn}
Joshua~C. Long et~al.
\newblock New experimental limits on macroscopic forces below 100- microns.
\newblock {\em Nature}, 421:922--925, 2003, hep-ph/0210004.

\bibitem{Wiseman:2001xt}
Toby Wiseman.
\newblock Relativistic stars in randall-sundrum gravity.
\newblock {\em Phys. Rev.}, D65:124007, 2002, hep-th/0111057.

\bibitem{Kudoh:2003xz}
Hideaki Kudoh, Takahiro Tanaka, and Takashi Nakamura.
\newblock Small localized black holes in braneworld: Formulation and numerical
  method.
\newblock {\em Phys. Rev.}, D68:024035, 2003, gr-qc/0301089.

\bibitem{Casadio:2002uv}
Roberto Casadio and Lorenzo Mazzacurati.
\newblock Bulk shape of brane-world black holes.
\newblock {\em Mod. Phys. Lett.}, A18:651--660, 2003, gr-qc/0205129.

\bibitem{Casadio:2001wh}
Roberto Casadio and Benjamin Harms.
\newblock Can black holes and naked singularities be detected in accelerators?
\newblock {\em Int. J. Mod. Phys.}, A17:4635--4646, 2002, hep-th/0110255.

\bibitem{Dadhich:2000am}
Naresh Dadhich, Roy Maartens, Philippos Papadopoulos, and Vahid Rezania.
\newblock Black holes on the brane.
\newblock {\em Phys. Lett.}, B487:1--6, 2000, hep-th/0003061.

\bibitem{Dadhich:2001ry}
Naresh Dadhich and S.~G. Ghosh.
\newblock Gravitational collapse of null fluid on the brane.
\newblock {\em Phys. Lett.}, B518:1--7, 2001, hep-th/0101019.

\bibitem{Bruni:2001fd}
Marco Bruni, Cristiano Germani, and Roy Maartens.
\newblock Gravitational collapse on the brane.
\newblock {\em Phys. Rev. Lett.}, 87:231302, 2001, gr-qc/0108013.

\bibitem{Govender:2001gq}
M~Govender and N~Dadhich.
\newblock Collapsing sphere on the brane radiates.
\newblock {\em Phys. Lett.}, B538:233--238, 2002, hep-th/0109086.

\bibitem{Aliev:2005bi}
A.~N. Aliev and A.~E. Gumrukcuoglu.
\newblock Charged rotating black holes on a 3-brane.
\newblock {\em Phys. Rev.}, D71:104027, 2005, hep-th/0502223.

\bibitem{Myers:1986un}
Robert~C. Myers and M.~J. Perry.
\newblock Black holes in higher dimensional space-times.
\newblock {\em Ann. Phys.}, 172:304, 1986.

\bibitem{Giddings:2001bu}
Steven~B. Giddings and Scott~D. Thomas.
\newblock High energy colliders as black hole factories: The end of short
  distance physics.
\newblock {\em Phys. Rev.}, D65:056010, 2002, hep-ph/0106219.

\bibitem{Emparan:2000rs}
Roberto Emparan, Gary~T. Horowitz, and Robert~C. Myers.
\newblock Black holes radiate mainly on the brane.
\newblock {\em Phys. Rev. Lett.}, 85:499--502, 2000, hep-th/0003118.

\bibitem{Harris:2003eg}
Chris~M. Harris and Panagiota Kanti.
\newblock Hawking radiation from a (4+n)-dimensional black hole: Exact results
  for the schwarzschild phase.
\newblock {\em JHEP}, 10:014, 2003, hep-ph/0309054.

\bibitem{Kanti:2004nr}
Panagiota Kanti.
\newblock Black holes in theories with large extra dimensions: A review.
\newblock {\em Int. J. Mod. Phys.}, A19:4899--4951, 2004, hep-ph/0402168.

\bibitem{Cardoso:2005vb}
Vitor Cardoso, Marco Cavaglia, and Leonardo Gualtieri.
\newblock Black hole particle emission in higher-dimensional spacetimes.
\newblock {\em Phys. Rev. Lett.}, 96:071301, 2006, hep-th/0512002.

\bibitem{Cardoso:2005mh}
Vitor Cardoso, Marco Cavaglia, and Leonardo Gualtieri.
\newblock Hawking emission of gravitons in higher dimensions: Non- rotating
  black holes.
\newblock {\em JHEP}, 02:021, 2006, hep-th/0512116.

\bibitem{Park:2005vw}
D.~K. Park.
\newblock Hawking radiation of the brane-localized graviton from a
  (4+n)-dimensional black hole.
\newblock 2005, hep-th/0512021.

\bibitem{Creek:2006ia}
S.~Creek, O.~Efthimiou, P.~Kanti, and K.~Tamvakis.
\newblock Graviton emission in the bulk from a higher-dimensional schwarzschild
  black hole.
\newblock {\em Phys. Lett.}, B635:39--49, 2006, hep-th/0601126.

\bibitem{1999ApJ...520..124G}
D.~E. {Gruber}, J.~L. {Matteson}, L.~E. {Peterson}, and G.~V. {Jung}.
\newblock {The Spectrum of Diffuse Cosmic Hard X-Rays Measured with HEAO 1}.
\newblock {\em \apj}, 520:124--129, July 1999.

\bibitem{2000AIPC..510..467W}
G.~{Weidenspointner}, M.~{Varendorff}, S.~C. {Kappadath}, K.~{Bennett},
  H.~{Bloemen}, R.~{Diehl}, W.~{Hermsen}, G.~G. {Lichti}, J.~{Ryan}, and
  V.~{Sch{\"o}felder}.
\newblock {The Cosmic Diffuse Gamma-Ray Background Measured with COMPTEL}.
\newblock In {\em American Institute of Physics Conference Series}, volume 510,
  pages 467--+, 2000.

\bibitem{Strong:2004ry}
A.~W. Strong, I.~V. Moskalenko, and O.~Reimer.
\newblock A new determination of the extragalactic diffuse gamma-ray background
  from egret data.
\newblock {\em Astrophys. J.}, 613:956--961, 2004, astro-ph/0405441.

\bibitem{Strong:2004de}
Andrew~W. Strong, Igor~V. Moskalenko, and Olaf Reimer.
\newblock Diffuse galactic continuum gamma rays. a model compatible with egret
  data and cosmic-ray measurements.
\newblock {\em Astrophys. J.}, 613:962--976, 2004, astro-ph/0406254.

\bibitem{Sreekumar:1997un}
P.~Sreekumar et~al.
\newblock Egret observations of the extragalactic gamma ray emission.
\newblock {\em Astrophys. J.}, 494:523--534, 1998, astro-ph/9709257.

\bibitem{Cornell:2005ux}
Alan~S. Cornell, Wade Naylor, and Misao Sasaki.
\newblock Graviton emission from a higher-dimensional black hole.
\newblock {\em JHEP}, 02:012, 2006, hep-th/0510009.

\bibitem{Kawasaki:2004yh}
Masahiro Kawasaki, Kazunori Kohri, and Takeo Moroi.
\newblock Hadronic decay of late-decaying particles and big-bang
  nucleosynthesis.
\newblock {\em Phys. Lett.}, B625:7--12, 2005, astro-ph/0402490.

\bibitem{Kawasaki:2004qu}
Masahiro Kawasaki, Kazunori Kohri, and Takeo Moroi.
\newblock Big-bang nucleosynthesis and hadronic decay of long-lived massive
  particles.
\newblock {\em Phys. Rev.}, D71:083502, 2005, astro-ph/0408426.

\bibitem{Kohri:2005wn}
Kazunori Kohri, Takeo Moroi, and Akira Yotsuyanagi.
\newblock Big-bang nucleosynthesis with unstable gravitino and upper bound on
  the reheating temperature.
\newblock 2005, hep-ph/0507245.

\bibitem{Kobayashi:2003cn}
Tsutomu Kobayashi, Hideaki Kudoh, and Takahiro Tanaka.
\newblock Primordial gravitational waves in inflationary braneworld.
\newblock {\em Phys. Rev.}, D68:044025, 2003, gr-qc/0305006.

\bibitem{Hiramatsu:2003iz}
Takashi Hiramatsu, Kazuya Koyama, and Atsushi Taruya.
\newblock Evolution of gravitational waves from inflationary brane- world:
  Numerical study of high-energy effects.
\newblock {\em Phys. Lett.}, B578:269--275, 2004, hep-th/0308072.

\bibitem{Hiramatsu:2004aa}
Takashi Hiramatsu, Kazuya Koyama, and Atsushi Taruya.
\newblock Evolution of gravitational waves in the high-energy regime of
  brane-world cosmology.
\newblock {\em Phys. Lett.}, B609:133--142, 2005, hep-th/0410247.

\bibitem{Kobayashi:2005jx}
Tsutomu Kobayashi and Takahiro Tanaka.
\newblock Quantum-mechanical generation of gravitational waves in braneworld.
\newblock {\em Phys. Rev.}, D71:124028, 2005, hep-th/0505065.

\bibitem{Kobayashi:2005dd}
Tsutomu Kobayashi and Takahiro Tanaka.
\newblock The spectrum of gravitational waves in randall-sundrum braneworld
  cosmology.
\newblock {\em Phys. Rev.}, D73:044005, 2006, hep-th/0511186.

\bibitem{Kinoshita:2005nx}
Shunichiro Kinoshita, Hideaki Kudoh, Yuuiti Sendouda, and Katsuhiko Sato.
\newblock Quadrupole formula for kaluza-klein modes in the braneworld.
\newblock {\em Class. Quant. Grav.}, 22:3911--3922, 2005, gr-qc/0505011.

\bibitem{Hiramatsu:2006bd}
Takashi Hiramatsu.
\newblock High-energy effects on the spectrum of inflationary gravitational
  wave background in braneworld cosmology.
\newblock {\em Phys. Rev.}, D73:084008, 2006, hep-th/0601105.

\bibitem{Kobayashi:2006pe}
Tsutomu Kobayashi.
\newblock Initial kaluza-klein fluctuations and inflationary gravitational
  waves in braneworld cosmology.
\newblock 2006, hep-th/0602168.

\bibitem{Seahra:2006tm}
Sanjeev~S. Seahra.
\newblock Gravitational waves and cosmological braneworlds: A characteristic
  evolution scheme.
\newblock 2006, hep-th/0602194.

\bibitem{Liddle:2003gw}
Andrew~R. Liddle and Anthony~J. Smith.
\newblock Observational constraints on braneworld chaotic inflation.
\newblock {\em Phys. Rev.}, D68:061301, 2003, astro-ph/0307017.

\bibitem{Tsujikawa:2003zd}
Shinji Tsujikawa and Andrew~R. Liddle.
\newblock Constraints on braneworld inflation from cmb anisotropies.
\newblock {\em JCAP}, 0403:001, 2004, astro-ph/0312162.

\end{thebibliography}

\end{document}